\documentclass[aip,graphicx,preprint]{revtex4-2}
\usepackage{mhchem, braket, graphicx, subcaption, tikz}
\usepackage{enumitem} 
\usepackage{comment}
\newcommand{\cm}{cm\ensuremath{^{-1}}}
\newcommand{\ansatz}{ex-\ensuremath{\lambda}}

\usepackage{float}
\usepackage{setspace}
\usepackage{multirow}
\usepackage{mathtools}

\begin{document}
	\title{Cumulants as the Variables of Density Cumulant Theory: A Path to Hermitian Triples}

\author{Jonathon P. Misiewicz}
\altaffiliation{
Also at Department of Chemistry and Cherry L. Emerson Center for Scientific Computation, Emory University, Atlanta, Georgia 30322, USA}
\author{Justin M. Turney}
\affiliation
{Center for Computational Quantum Chemistry, University of Georgia, Athens, Georgia 30602, United States}
\author{Henry F. Schaefer III}
\email{ccq@uga.edu}
\affiliation
{Center for Computational Quantum Chemistry, University of Georgia, Athens, Georgia 30602, United States}

	\date{\today}

\begin{abstract}
We study the combination of orbital-optimized density cumulant theory and a new parameterization of the reduced density matrices in which the variables are the particle-hole cumulant elements. We call this combination O$\lambda$DCT. We find that this new ansatz solves problems identified in the previous unitary coupled cluster ansatz for density cumulant theory: the theory is now free of near-zero denominators between occupied and virtual blocks, can correctly describe the dissociation of \ce{H2}, and is rigorously size-extensive. In addition, the new ansatz has fewer terms than the previous unitary ansatz, and the optimal orbitals delivered by the exact theory are the natural orbitals. Numerical studies on systems amenable to full configuration interaction show that the amplitudes from the previous ODC-12 method approximate the exact amplitudes predicted by this ansatz. Studies on equilibrium properties of diatomic molecules show that even with the new ansatz, it is necessary to include triples to improve the accuracy of the method compared to orbital optimized linearized coupled cluster doubles. With a simple iterative triples correction, O$\lambda$DCT outperforms other orbital-optimized methods truncated at comparable levels in the amplitudes, as well as CCSD(T). By adding four more terms to the cumulant parameterization, O$\lambda$DCT outperforms CCSDT while having the same $\mathcal{O}(V^5 O^3)$ scaling.
\end{abstract}

	\maketitle

\section{Introduction}

Density cumulant theory\cite{Kutzelnigg:2006p171101} (DCT) is a family of electronic structure methods in which the electronic energy is computed by parameterizing the cumulant\cite{Misiewicz:2020pX, Hanauer:2012p50, Mazziotti:2011p244, Mazziotti:1998p419, Kutzelnigg:1999p2800, Ziesche:2000p33, Piris:2014p1169, Torre:2003p127} of the 2-electron reduced density matrix (2RDM),\cite{Ciosloswki2000, ACP134} from which the 1-electron reduced density matrix (1RDM) is then constructed.\cite{Kutzelnigg:2006p171101, Nooijen:1999p8356, Herbert_2007} Any parameterization of the 2RDM cumulant defines a method of this family, and an exact parameterization produces an exact method. 

The development of new DCT parameterizations is chiefly motivated\cite{Sokolov:2014p074111, Mullinax_2015, Cioslowski:2019p4862, Misiewicz:2020p244102} by the success of the orbital-optimized ODC-12 method.\cite{Sokolov:2013p024107, Sokolov:2013p204110} The ODC-12 parameters consist of orbital rotation amplitudes\cite{Kats:2014p061101, Sherrill:1998p4171, Bozkaya:2011p104103, Sokolov:2013p204110, Kats:2018p13, Stein:2014p214113, Robinson:2011p044113, Bozkaya:2014p204105, Pavo_evi_:2020p1578, Bozkaya:2014p2371, Lee2018} and rank two excitation-type amplitudes, analogous to those of orbital-optimized coupled cluster doubles.\cite{Bozkaya:2011p104103} The resulting ODC-12 method boasts superior accuracy to CCSD for the same computational scaling,\cite{Copan:2014p2389, Sokolov:2013p204110, Wang:2016p4833} more tolerance of strong correlation than CCSD,\cite{Mullinax_2015} a straightforward analytic gradient theory,\cite{Sokolov:2013p204110} and a response theory that is both efficient and hermitian.\cite{Copan:2018p4097, Peng:2019p1840} We intend to extend this method to a hierarchy of increasingly accurate DCT methods.

Just as there are different wavefunction parameterizations, there are different cumulant parameterizations. It is not \textit{a priori} obvious which cumulant parameterization one should choose to continue DCT. 
Kutzenigg's original manuscript remarked that ``there is some flexibility in this step'' and therefore ``experience ought to be gained'' as to the appropriate parameterization.\cite{Kutzelnigg:2006p171101} In that article, all elements of the 2RDM cumulant were written as functions of the particle-hole 2RDM cumulant elements. We are not aware of any proposed cumulant parameterization for DCT that maintains this property, only an ambiguous remark.\cite{Kutzelnigg2012}

Instead, Kutzelnigg and others\cite{Sokolov:2014p074111} parameterized the DCT cumulant by unitary coupled cluster\cite{Cooper:2010p234102, Evangelista:2011p224102, Chen:2012p014108, Bartlett:1989p133, Taube:2006p3393, Kutzelnigg:1991p349, Kutzelnigg_2010, Kutzelnigg_1977, Szalay:1995p281, Misiewicz2021:I} amplitudes, for historical reasons\cite{Kutzelnigg:2004p7350}. The resulting orbital-optimized, unitary ansatz\cite{Sokolov:2014p074111} is called OUDCT. The experience we have gained is that this parameterization has undesirable features:\cite{Misiewicz:2020p244102}

\begin{enumerate}
	\item When the theory uses optimal orbitals and zeroes singles amplitudes, the 1RDM will in general not be block-diagonal in the occupied and virtual blocks. This feature is inherited from unitary coupled cluster and causes two complications:\cite{Misiewicz:2020p244102}
	\begin{enumerate}
		\item The formal accuracy estimates are less optimistic than those of unitary coupled cluster. To construct a 1RDM correct to degree $n$ as a Taylor series in the unitary coupled cluster amplitudes, DCT requires the cumulant to be correct to degree $n+2$ in those amplitudes.
		\item The amplitude residual expressions of any DCT ansatz contain terms of the form
		\begin{equation*}
			\frac{1}{n_{p^\prime} + n_{q^\prime} - 1} \frac{\partial}{\partial t} d^{p^\prime}_{q^\prime}
		\end{equation*}
		where $n_{p^\prime}$ is the occupation number of natural spinorbital $p$,\cite{Davidson_1972}
		and $d$ is the partial trace of the 2RDM cumulant.\cite{Sokolov:2013p024107} The denominators will tend to zero when $p^\prime$ is highly occupied and $q^\prime$ is highly virtual, which occurs even for weak correlation. If these terms are not removed, they may produce numerical instability. The terms can be removed if $d$ can be guaranteed block-diagonal, as then $\frac{\partial}{\partial t} d^{p^\prime}_{q^\prime} = 0$ for occupied and virtual natural spinorbitals. But in the general case, OUDCT cannot remove the terms and therefore is not numerically stable.
	\end{enumerate}
	\noindent While using natural spinorbitals instead of optimal orbitals would resolve these problems, it sacrifices the simplicity of the analytic gradients: the energy functional is no longer variational, so it will be necessary to consider both orbital and amplitude responses,\cite{Sokolov:2012p054105} and differentiate between the RDMs used in the energy computation and the response RDMs with these extra response terms. With optimal orbitals, there are no response terms.\cite{Sokolov:2013p204110} Further, the corresponding unitary singles amplitudes may need to be explicitly included for the method to be formally exact, as they are no longer zero when starting from the natural determinant.\cite{Misiewicz:2020p244102} The sheer number of tensor contractions involving singles discourages their inclusion.
	\item Methods obtained by a Taylor series truncation of the exact OUDCT ansatz produce \ce{H2} dissociation curves that do not improve across the entire curve as terms of increasing degree are included. Compared to a degree two truncation, truncating at degrees three, four, or five in the OUDCT amplitudes give worse results far from equilibrium, with some methods not converging. As \ce{H2} is an especially simple bond dissociation problem, the truncated OUDCT ansatz is not expected to perform reliably for bond dissociation or systems with emergent strong correlation. This important feature of ODC-12 has been lost,\cite{Misiewicz:2020p244102} and it is not obvious why ODC-12 had it or how OUDCT sacrificed it.
	\item More recently, we have discovered OUDCT methods that are not size-extensive. This feature is inherited from a class of diagrams in the underlying unitary coupled cluster ansatz.\cite{Szalay:1995p281} We shall discuss this in Section \ref{subsec:oudct-size-extensivity}.
\end{enumerate}

In this article, we resolve all three of these problems in a single stroke: replace the unitary coupled cluster ansatz with a new ansatz, which we call \ansatz, in which the parameters are the particle-hole cumulants themselves. In contrast to past work in electronic structure theories with cumulants,\cite{Mazziotti_2010, Sokolov:2014p074111} the variables are not merely a ``first-order'' contribution to the particle-hole cumulants that will disagree with the particle-hole cumulant elements in the exact theory. Our variables are the particle-hole cumulants exactly. This ansatz has not been previously reported, and we will defer its general formulation to future research,\cite{Misiewicz2021:I} as the only derivation we have found requires new and elaborate theoretical machinery.

After a brief review of background in Section \ref{sec:bg}, we define $\lambda$DCT, which is DCT using \ansatz. We combine this with orbital optimization to produce O$\lambda$DCT. Once the method is defined, Section \ref{sec:formal} considers formal properties of the ansatz, emphasizing how the \ansatz\ ansatz naturally emerges as a way to avoid the problems of OUDCT discussed above. In particular, we consider:

\begin{enumerate}
	\item Do near-zero denominators appear in O$\lambda$DCT theories? If so, how has the problem of near-zero denominators changed from the OUDCT ansatz, and what can be done to control the problem further?
	\item For an analytically solvable toy model of \ce{H2} dissociation, can we explain the poor performance of truncated OUDCT methods and show how O$\lambda$DCT methods perform better?
	\item Are all methods of the ansatz size-extensive?
	\item How does the number of terms in O$\lambda$DCT theories compare to the number of terms in OUDCT?
\end{enumerate}

On the numerical side, we have two primary objectives. Our first is to affirm the \textit{correctness} of our ansatz by showing that the optimized variables of our approximate O$\lambda$DCT methods truly do approximate the cumulant. Applied to our new methods, this supports the correctness of the terms we compute in the exact theory. Applied to ODC-12, this justifies interpreting ODC-12 via the \ansatz\ ansatz, as opposed to some other ansatz. Second, we wish to evaluate the \textit{usefulness} of our ansatz by assessing how competitive O$\lambda$DCT methods are with conventional coupled cluster methods and alternative orbital-optimized parameterizations of the reduced density matrices. In Section \ref{sec:numerical}, we consider:
\begin{enumerate}
	\item Do truncations of the O$\lambda$DCT ansatz improve on \ce{H2} dissociation for a non-minimal basis set?
	\item Do the amplitudes of approximate O$\lambda$DCT, including ODC-12, follow those of the exact ansatz?
	\item Can computing more terms restricted to rank two amplitudes improve the description of equilibrium properties for systems of more than two electrons?
	\item How does a simple inclusion of iterative triples, in all the various orbital-optimized RDM parameterization methods we consider, affect the description of equilibrium properties and bond dissociation for systems of more than two electrons?
	\item How does the accuracy of our approximate O$\lambda$DCT ansatz compare with existing CCSD(T) and CCSDT ans{\"a}tze?
\end{enumerate}

In particular, with a natural parameterization involving triples within O$\lambda$DCT, one arrives at an O$\lambda$DCT method \textit{more accurate than CCSDT} but with the same $\mathcal{O}(V^5 O^3)$ scaling of CCSDT. Further, it is of comparable accuracy to other reduced density matrix parameterizations at the same degree of truncation, but with far fewer terms.

\section{Theoretical Background}
\label{sec:bg}

\subsection{Abstract Density Cumulant Theory}
\label{subsec:abstract}
This subsection briefly repeats the most salient points of DCT, abstracting away the cumulant parameterization. For expanded discussion, see Section IIA of our study of OUDCT.\cite{Misiewicz:2020p244102}

In DCT, the electronic energy is written as
\begin{equation}
	\label{eq:cumulant-energy}
	E = (h_p^q + \frac{1}{2} \bar{g}_{pr}^{qs} \gamma_s^r)  \gamma^p_q + \frac{1}{4} \bar{g}_{pq}^{rs} \lambda^{pq}_{rs}
\end{equation}

\noindent where $\gamma$ is the 1-electron reduced density matrix (1RDM)

\begin{equation}
	\label{eq:1rdm}
	\gamma^p_q = \frac{\bra{\Psi} a^p_q \ket{\Psi}}{\braket{\Psi | \Psi}}
\end{equation}

\noindent and $\lambda$ is the size-extensive part, the cumulant,\cite{Misiewicz:2020pX, Mazziotti:1998p419, Kutzelnigg:1999p2800} of the 2-electron reduced density matrix,

\begin{equation}
	\label{eq:2rdm}
	\gamma^{pq}_{rs} = \frac{\bra{\Psi} a^{pq}_{rs} \ket{\Psi}}{\braket{\Psi | \Psi}}
\end{equation}

\noindent and

\begin{equation}
	\label{eq:2rdm-decomp}
	\gamma^{pq}_{rs} = \lambda^{pq}_{rs} + \gamma^p_r\gamma^q_s - \gamma^p_s \gamma^q_r \quad .
\end{equation}

\noindent We also define $h_p^q$ as the standard one-electron integral, $\bra{\phi_p} \hat{h} \ket{\phi_q}$, and $\bar{g}_{pq}^{rs}$ as the antisymmetrized electron repulsion integral, $\bra{pq} \ket{rs}$. We use the notation introduced by Kutzelnigg in Reference \citenum{Kutzelnigg:1982p3081} for writing the vacuum-normal, particle-conserving second quantized fermionic operators ($a^p_q = a_p^\dagger a_q$ and $a^{pq}_{rs} = a_p^\dagger a_q^\dagger a_s a_r$). We also use the Einstein summation convention throughout this article: summation is implied over all indices that appear twice in a product of tensors. The indices $p, q, r, s$ label a general orbital. The indices $i, j, k$ label an occupied orbital, and the indices $a, b, c$ label a virtual orbital.

The energy function \eqref{eq:cumulant-energy} applies only to a pair of $\gamma$ and $\lambda$ that derives from some normalized wavefunction, $\ket{\Psi}$. Such a pair of $\gamma$ and $\lambda$ is said to be pure $n$-representable.\cite{Coleman:1963p668, Schilling:2018p231102, Klyachko:2006p72, Altunbulak:2008p287, Coleman:1978p67, Mazziotti:2016p032516} To compute an energy, the pure $n$-representable $\gamma$ and $\lambda$ are parameterized, and the energy function is variationally minimized with respect to those parameters. This general strategy of parameterizing cumulants and minimizing the induced energy function has been employed by other researchers.\cite{Mazziotti:2011p244, Mazziotti_2010, Hollett:2020p014101, Hollett:2016p084106, Piris:2014p1169}

While an explicit parameterization of $\lambda$ is necessary, an explicit parameterization of $\gamma$ is not. It may instead be derived from the parameterization of $\lambda$. This implicit parameterization is the distinctive feature of DCT. Defining the cumulant partial trace\cite{Kutzelnigg:2006p171101, Mazziotti_2010} $d$ with

\begin{equation}
	\label{eq:def-d}
	d^p_q = \lambda^{pr}_{qr}
\end{equation}

\noindent it follows that

\begin{equation}
	\label{eq:rdmc-partial-trace}
	d^p_q = (\gamma^2 - \gamma)^p_q \quad .
\end{equation}

\noindent While multiple possible 1RDMs satisfy this equation for a given $d$, the solutions are exactly the matrices obtainable by the following procedure: choose an eigenbasis of $d$, and let eigenvector $p^\prime$ with $d$ eigenvalue $d_{p^\prime}$ be an eigenvector of $\gamma$ with eigenvalue \begin{equation}
	\label{eq:d-to-gamma-eval}
	\gamma_{p^\prime} = \frac{1 \pm \sqrt{1+4 d_{p^\prime}}}{2} \quad .
\end{equation}

\noindent The choice of the plus or minus sign determines whether the orbital occupation number is above 0.5 (occupied-like) or below 0.5 (virtual-like).\cite{Misiewicz:2020p244102}

When $\gamma$ is constructed from $\lambda$ in this way, the derivative of the energy with respect to amplitude $t$ is given\cite{Sokolov:2013p024107} by

\begin{equation}
	\label{eq:amplitude-gradient}
	\frac{\partial E}{\partial t} = \tilde{F}_p^q \frac{\partial d^p_q}{\partial t} + \bar{g}_{pq}^{rs} \frac{\partial \lambda^{pq}_{rs}}{\partial t}
\end{equation}

\noindent defining

\begin{equation}
	\label{eq:gen-fock}
	\tilde{F}_{p^\prime}^{q^\prime} = \frac{h_{p^\prime}^{q^\prime} + \bar{g}_{p^\prime r^\prime}^{q^\prime s^\prime} \gamma^{r^\prime}_{s^\prime}}{n_{p^\prime} + n_{q^\prime} - 1}
\end{equation}

\noindent where $n_{p^\prime}$ is the occupation number of natural spinorbital $p^\prime$. Equation \eqref{eq:gen-fock} contains denominators that may be near zero if the sum of two natural orbital occupation numbers approximates 1. These terms vanish if the associated $\frac{\partial}{\partial t} d^p_q$ is zero, which depends on the cumulant parameterization appearing through \eqref{eq:def-d}. We emphasize that \eqref{eq:gen-fock} must be first computed in the basis of natural spin-orbitals and then transformed back to whatever orbital basis is used in \eqref{eq:amplitude-gradient}.

To define an approximate DCT method, it remains to parameterize $\lambda$. If $\lambda$ is parameterized exactly, all we have done is rearrange an exact parameterization of the 1RDM and 2RDM: we will recover the same energy for the same parameters. If $\lambda$ is parameterized only approximately, then DCT couples the approximation used for the 1RDM and the product of 1RDMs to the approximation used for the 2RDM cumulant. DCT does not specify how $\lambda$ should be parameterized, although it makes some choices more convenient.

\subsection{Cumulant Parameterizations}

We now survey the four parameterizations of the reduced density matrices and their cumulants that we shall combine with DCT in this study. These four are, to the authors' knowledge, the only known RDM parameterizations with one wavefunction's worth of parameters and a connected expansion of the energy. Approximating any RDM parameterization induces an energy function, and variationally minimizing that induces a method to compute ground-state electronic energies. By feeding an approximate cumulant of these parameterizations into DCT as described into the last section, these produce an approximate DCT method.

Our first two parameterizations originate from a wavefunction parameterization. From this, we immediately have a parameterization of the reduced density matrices. From the formula for the 2RDM cumualnt, \eqref{eq:2rdm-decomp}, we immediately have a parameterization of the cumulants. If the parameters are additively separable, the cumulant consists of the connected tensors of the RDM, by an argument\cite{Misiewicz:2020p244102} reminiscent of generalized extensivity.\cite{Hanrath2009, Nooijen2005}

In the unitary coupled cluster (UCC) parameterization of $\ket{\Psi}$, we employ

\begin{equation}
	\label{eq:ucc-wfn}
	\ket{\Psi} = \exp(\hat{\sigma} - \hat{\sigma}^\dagger) \ket{\Phi}
\end{equation}

\noindent where

\begin{equation}
	\hat{\sigma} = \hat{\sigma}_1 + \hat{\sigma}_2 + ...
\end{equation}

\noindent and

\begin{align}
	\hat{\sigma}_1 =& \frac{1}{(1!)^2} \sigma^i_a a^a_i \\%
	\hat{\sigma}_2 =& \frac{1}{(2!)^2} \sigma^{ij}_{ab} a^{ab}_{ij} \\
	\hat{\sigma}_3 =& \frac{1}{(3!)^2} \sigma^{ijk}_{abc} a^{abc}_{ijk}
\end{align}

\noindent and so forth. This immediately writes the reduced density matrices as functions of $\sigma$ by inserting \eqref{eq:ucc-wfn} into the definition of the 2RDM, \eqref{eq:2rdm}. We denote this parameterization VUCC, to emphasize that \textit{unitary coupled cluster} is used in a \textit{variational} rather than projective sense.

The variational coupled cluster (VCC) parameterization is similarly derived from the wavefunction parameterization

\begin{equation}
	\label{eq:cc}
	\ket{\Psi} = \exp(T) \ket{\Phi}
\end{equation}

\noindent where

\begin{equation}
	\label{eq:cc-t}
	T = T_1 + T_2 + ...
\end{equation}

\noindent and

\begin{align}
	T_1 =& \frac{1}{(1!)^2} t^i_a a^a_i \\
	T_2 =& \frac{1}{(2!)^2} t^{ij}_{ab} a^{ab}_{ij} \\
	T_3 =& \frac{1}{(3!)^2} t^{ijk}_{abc} a^{abc}_{ijk} \quad .
\end{align}

Our two remaining parameterizations \textit{do not} start with a wavefunction parameterization. Strongly-connected variational coupled cluster\cite{Szalay:1995p281} (SCVCC) instead starts from the explicit form of the expectation value of operator $O$ within VCC, given by
\begin{align}
	\langle O \rangle &= \frac{\braket{\Phi | \exp(T^\dagger) O \exp(T) | s\Phi}}{\braket{\Phi | \exp(T^\dagger) \exp(T) | \Phi}} \\
	& = \braket{\Phi | \exp(T^\dagger) O \exp(T) | \Phi}_C \label{eq:vcc-rdm}
\end{align}

\noindent where the subscript $C$ denotes that the only tensors retained are those consisting only of connected diagrams, or connected contractions within Wick's Theorem.\cite{Cizek1969, Pal:1982p261, Shavitt:2009}

Now, \eqref{eq:vcc-rdm} consists of some terms that will remain connected upon removal of any amplitude, and some terms that will not. The first class of terms are called storngly connected, and the second class of terms is called weakly connected. For reasons we shall discuss in Section \ref{subsec:oudct-size-extensivity}, Szalay, Nooijen, and Bartlett sought to remove the weakly connected terms from \eqref{eq:vcc-rdm}. They defined a new cluster operator $T^\ast$ from the VCC cluster operator $T$ and claimed that

\begin{align}
	\langle O \rangle &= \braket{\Phi | \exp((T^\ast)^\dagger) O \exp(T^\ast) | \Phi}_\textsubscript{SC} \label{eq:scvcc-rdm}
\end{align}

\noindent expresses the expectation value of operator $O$, \eqref{eq:vcc-rdm}, using the amplitudes of $T^\ast$. The subscript SC denotes a restriction to the strongly connected terms. The formula \eqref{eq:scvcc-rdm} defines SCVCC.

Two cautions are in order. First, Reference \citenum{Szalay:1995p281} was concerned with the special case that $O = H$, but their argument is just as valid for general $O$, so if their argument holds, they have an $n$-representable parameterization of the reduced density matrices, within radius of convergence. Second, their justification of \eqref{eq:scvcc-rdm} relies on a substitution operation on diagrams that, in the judgment of the present authors, requires further mathematical justification. We shall pursue this in future research.\cite{Misiewicz2021:I} At present, it is enough to know that their conclusion is correct.

Our last ansatz is \ansatz, where the variables are the particle-hole cumulants. This ansatz has not, to our knowledge, been described in any previous literature. In this article, we sketch a tedious but simple derivation of the ansatz, leaving elaboration and a less tedious rule to calculate weight factors to future research.\cite{Misiewicz2021:I} Section 1 of our Supporting Information explicitly performs this tedious derivation to the orders we implement here.

Observe that in SCVCC, if we define the parameters to be $t^\ast$, we know

\begin{equation}
	\lambda^{ij\cdots}_{ab\cdots} = (t^\ast)^{ij\cdots}_{ab\cdots} + \mathcal{O}((t^\ast)^2)
\end{equation}

\noindent and therefore

\begin{equation}
	\label{eq:t-flip}
	(t^\ast)^{ij\cdots}_{ab\cdots} = \lambda^{ij\cdots}_{ab\cdots} - \mathcal{O}((t^\ast)^2) \quad .
\end{equation}

\noindent To write a cumulant element as a function of the particle-hole cumulants, $\lambda$, take the SCVCC parameterization of that element and recursively insert \eqref{eq:t-flip} whenever a $t$ occurs. Each insertion produces two kinds of terms: one where a $t$ has transformed to a $\lambda$, and one that has increased the total number of amplitudes. After enough substitutions, all terms must have either gone to more than $n$ amplitudes or had all amplitudes $t$ converted to $\lambda$. At that point, the cumulant element has been expressed as a function of the particle-hole cumulants to degree $n$, and we have the derived terms of the ansatz to that degree.

We note that this derivation can be done just as well from any other parameterization of the reduced density matrices and their cumulants, but starting from SCVCC will be convenient for Section \ref{subsec:oudct-size-extensivity}.

\subsection{Orbital Optimization}

All these ans{\"a}tze contain rank-one amplitudes, indexed by an occupied and virtual orbital. We may choose not to vary these parameters, but to instead vary the orbitals. The orbital rotation parameters are also indexed by one occupied and one virtual orbital,\cite{Sherrill:1998p4171, Bozkaya:2011p104103} so we have preserved the number of parameters. This is known as orbital optimization. In this study, we apply orbital optimization to all of our parameterizations, whether using DCT or not. By combining DCT, the \ansatz\ ansatz, and orbital optimization, we define what we call O$\lambda$DCT.

Orbital optimization is known to correct for spin contamination and spatial symmetry breaking in reference states, and has been found especially helpful for describing difficult species, such as open-shell species and transition states. (See Reference \citenum{Bozkaya2020} and references therein.)

In order for the ans{\"a}tze to be exact, there must exist orbitals such that the exact ground-state wavefunction has the rank one parameters be zero for all occupied and virtual orbitals. For the VCC ansatz, these orbitals are guaranteed to exist and are called the Brueckner orbitals.\cite{Brueckner:1956p1008, Nesbet:1958p1632, Shavitt:2009, osti_4771839} For the \ansatz\ ansatz, the natural orbitals,\cite{Davidson_1972} the eigenfunctions of the 1RDM, certainly have the property $\lambda^i_a = 0$. Any orbitals that have $\lambda^i_a = 0$ must be related to the natural orbitals by independent similarity transformations of the occupied and virtual spaces. Therefore, up to separate diagonalization of the occupied and virtual spaces, the optimal orbitals of \ansatz\ are the natural orbitals.

It is known that the optimal unitary orbitals cannot in general be natural orbitals.\cite{Misiewicz:2020p244102} Formally, this is a consequence of the VUCC terms of $\gamma^i_a$ that do not contain rank one amplitudes, e.g., $\frac{1}{6} \sigma^{jk}_{ab} \sigma^{il}_{cd} \sigma_{jkl}^ {bcd}$. The same tensor contraction shows that SCVCC's optimal orbitals cannot in general be the natural orbitals, either.

\section{Formal Analysis}
\label{sec:formal}

In the previous section, we defined O$\lambda$DCT. We have not yet motivated \ansatz\ as the source of the cumulant parameterization of DCT.

Section \ref{sec:why} shall explain how O$\lambda$DCT satisfies each of the formal desiderata from the introduction in turn. Each of these requirements puts a strict constraint on the parameterizations we can consider. While we cannot conclude that O$\lambda$DCT is the unique theory satisfying these requirements, careful consideration of them naturally leads to the particle-hole cumulants as our parameters of choice. In Section \ref{sec:explicit}, we show explicit formulas for terms of the cumulant in O$\lambda$DCT and compare these to terms from the OUDCT theory.\cite{Sokolov:2014p074111}

\subsection{Why \ansatz?}
\label{sec:why}

\subsubsection{Block Diagonal Structure of $d$ and $\gamma$}
\label{subsec:block-diag}

As explained in the introduction, to avoid singularities, we want to enforce that the occupied-virtual block of the 1RDM, $\gamma^i_a$, is zero. We also want to use optimized orbitals, which requires that our singles parameters, $p^i_a$, are zero. If the occupied-virtual block of the 1RDM is given by the parameters, $p^i_a$, we can have both results at once. If we want to extend the rank one parameters to a family of additively separable parameters, the natural choice is the size-extensive part of the particle-hole RDM elements, i.e., the \ansatz\ ansatz. In contrast, VCC, SCVCC, or VUCC all have terms in the parameterization of $\gamma^i_a$ that do not vanish when their respective $p^i_a$ are zero and therefore cannot be the parameterizations we seek.

There is a slight wrinkle in our logic: DCT determines the 1RDM from diagonalization of $d$ by \eqref{eq:d-to-gamma-eval}. To establish that setting the parameters $\gamma^i_a$ to zero makes the corresponding DCT RDM elements zero, we must prove that setting $\gamma^i_a$ to zero makes the off-diagonal blocks of $d$ zero.

To prove this, we use the equation relating $d$ to the $\gamma$ in any exact ansatz, \eqref{eq:rdmc-partial-trace}. The occupied-virtual block of \eqref{eq:rdmc-partial-trace} is

\begin{equation}
	d^i_a = \gamma^i_j \gamma^j_a + \gamma^i_b \gamma^b_a - \gamma^i_a \quad .
\end{equation}

\noindent When all $\gamma^i_a$ are zero, the right-hand side is zero. Therefore, every term in the power series expansion of $d^i_a$ must be at least degree one in the rank-one amplitudes. When the cumulant is computed to a particular degree in the Taylor series, the resulting partial trace must equal the expansion of $d$ to the same degree, so as long as $\gamma^i_a = 0$, $d$ is block-diagonal, as desired.

This resolves the first problem mentioned in the introduction: O$\lambda$DCT is guaranteed to avoid near-zero denominators between occupied and virtual orbitals as long as no degree of the Taylor series is computed only partially. We make two more technical observations:

\begin{enumerate}
	
	\item We can make better estimates of the 1RDM from a given 2RDM cumulant truncation. In the OUDCT ansatz, the 2RDM cumulant is needed to degree $n+2$ to get the 1RDM correct to degree $n$. In the O$\lambda$DCT ansatz, the 2RDM cumulant is only needed to degree $n$ by the following argument: Equation \eqref{eq:rdmc-partial-trace} separates into one equation for the occupied block and one equation for the virtual block. Separate the 1RDM into a constant term, $\kappa$, which is the identity for the occupied block and zero for the virtual block, and the non-constant terms, $\tau$. Then
	\begin{equation}
		\label{eq:do}
		\tau^i_j = + d_o - d_o^2 + 2d_o^3 - 5 d_o^4 + ... + (-1)^{n-1} C_{n-1} d_o^n + ...
	\end{equation}
	
	\noindent and
	
	\begin{equation}
		\label{eq:dv}
		\tau^a_b = - d_v + d_v^2 - 2d_v^3 + 5 d_v^4 - ... - (-1)^{n-1} C_{n-1} d_v^n + ...
	\end{equation}
	
	\noindent where $d_o$ refers to the occupied block of $d$ in \eqref{eq:do} and $d_v$ refers tothe virtual block in \eqref{eq:dv}. $C_n$ is the $n$th Catalan number. The Catalan numbers are a sequence of numbers that appear in numerous combinatorial problems and the generating function of the scalar analogue of \eqref{eq:d-to-gamma-eval}.\cite{Stanley:1999} To the skeptical reader, these findings are consistent with our previous demonstration of $n+2$ dependence for the OUDCT ansatz, found in the Supporting Information of Reference \citenum{Misiewicz:2020p244102}. The $n+2$ dependence was shown in the occupied-virtual block, which is $0$ here. The occupied-occupied and virtual-virtual blocks both had $n$ dependence.
	\item All terms of \eqref{eq:amplitude-gradient} involving a natural orbital from the occupied block and a natural orbital from the virtual block are eliminated. In practice, this should agree with choosing the occupied orbitals to be those with greatest occupation numbers, but this is not logically necessary. In the single-reference setting, this should resolve the singularities, but if natural orbital occupation number $n_{p^\prime} \approx 0.5$, then $(2 n_{p^\prime} - 1) \approx 0$, and the denominator \eqref{eq:gen-fock} becomes near singular. No change of variables can remedy the situation, as the derivative $\frac{\partial d^p_p}{\partial t}$ being zero for a region would mean that the natural orbital occupation number was constant there. O$\lambda$DCT ameliorates the singularities as much as any change of variables can.
\end{enumerate}

\subsubsection{\ce{H2} Dissociation}
\label{sec:formal-h2}

Our second desideratum for a new DCT ansatz is that approximating our new ansatz should allow DCT to describe \ce{H2} dissociation at least as well as ODC-12 does. To explain why \ansatz\ is a natural solution to this problem, we must first identify precisely why OUDCT had so much difficulty with this system, when VUCC did not.

Consider the $^1 \Sigma_g^+$ ground state of \ce{H2} with a STO-3G basis set. By symmetry arguments, only two configurations can contribute to the ground-state wavefunction: the $\sigma_g^2$ bonding orbitals and the $\sigma_u^2$ antibonding orbitals. Therefore, we only need one parameter to represent the reduced density matrices, and there are no symmetry-allowed orbital rotations. This allows us to study the problem analytically.

\begin{figure}
	\caption{Energy of \ce{H2} at 3.2 bohr with the STO-3G basis set, as predicted by various OUDCT approximations. Due to the minimal basis set, there is no orbital optimization problem. Note that the degree two approximation predicts the correct minimum energy, but at a different parameter compared to the exact theory. The dashed red line indicates the full configuration interaction solution.}
	\includegraphics[width=0.8\linewidth]{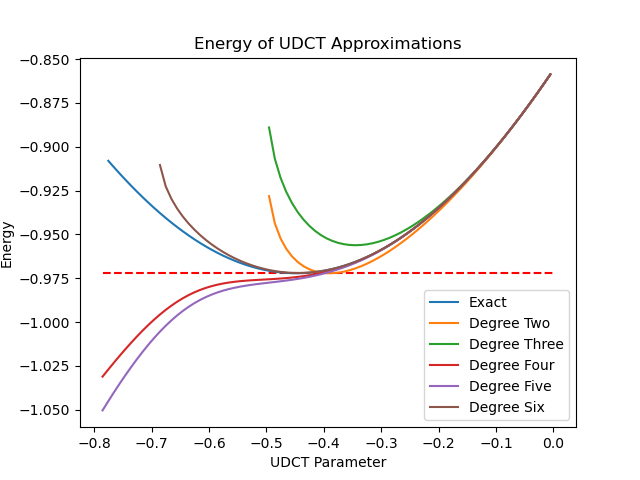}
	\label{fig:h2-toy-energy}
\end{figure}

Figure \ref{fig:h2-toy-energy} shows the energy as a function of that one parameter for various OUDCT truncations. We make three observations:

\begin{enumerate}
	\item The degree three theory has a large increase in energy compared to ODC-12, and the degree four and five theories do not possess a minimum. This agrees qualitatively with our cc-pVDZ results.\cite{Misiewicz:2020p244102}
	\item The minimum-energy parameter of ODC-12 is quite different than the value of the exact VUCC ansatz. This agrees qualitatively with our previous cc-pVDZ results.\cite{Misiewicz:2020p244102}
	\item The minimum energy of ODC-12 the same as full configuration interaction. This was not true within cc-pVDZ.\cite{Misiewicz:2020p244102}
\end{enumerate}

\begin{figure}
	\caption{The lone non-zero value of the cumulant partial trace, $d$, of \ce{H2} at 3.2 bohr with the STO-3G basis set, as predicted by various OUDCT approximations. $d$ is an even function.}
	\includegraphics[width=0.8\linewidth]{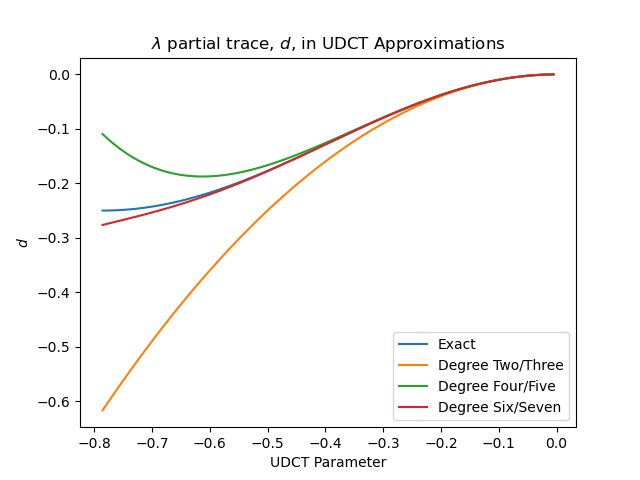}
	\label{fig:h2-toy-d}
\end{figure}

Let us first consider why the higher degree truncations fail. Probing the various terms in the energy, we find that all energy contributions involving the 1RDM have poor accuracy, as judged by VUCC. Recall that in DCT, the 1RDM is constructed from the matrix $d$, according to \eqref{eq:d-to-gamma-eval}. For this system, there is only one non-zero value of the matrix $d$, and its value as a function of the parameter is plotted in Figure \ref{fig:h2-toy-d}.

As a function of the UCC parameter, $d$ is near-flat in the neighborhoods of $0$ (where the Taylor expansion is centered) and at $- \frac{\pi}{4}$ (where $d = -0.25$, which is a critical point for reconstruction of $\gamma$). There is an inflection point between the two, and capturing this even qualitatively requires a \textit{sixth order} polynomial fit. The degree four and five approximations have an upswing in $d$ for $ t > -0.43$, and this leads to the error in the final energy. This suggests that for DCT to be accurate, we require a parameterization in which the expansion of $d$ is rapidly convergent.

We now return to the ODC-12 parameterization. It is known that for this toy system, when the cumulant elements are parameterized in terms of the particle-hole cumulants, the parameterization terminates at degree two, and one arrives at the cumulant parameterization of ODC-12.\cite{Kutzelnigg2012} In other words: \textit{for this toy system, O$\lambda$DCT is already exact at degree two, and OUDCT at degree two has the exact minimum energy because it is coincident with O$\lambda$DCT at this low degree}.

By this argument, the particle-hole cumulants are the natural choice of rank two variables for DCT. To extend this to a family of size-extensive variables, once again, the natural choice is to make the particle-hole cumulants the variables at all ranks. We expect that this choice means O$\lambda$DCT allows for more rapid convergence of $d$, increasing the accuracy of the theory.

\subsubsection{Size-Extensivity}
\label{subsec:oudct-size-extensivity}

During the course of our numerical studies, we discovered OUDCT parameterizations that were not size-extensive, despite claims that DCT is size-extensive even when approximations are introduced.\cite{Kutzelnigg:2006p171101, Copan:2018p4097, Sokolov:2014p074111, Wang:2016p4833, Mullinax_2015, Copan:2014p2389, Sokolov:2013p024107, Sokolov:2013p204110, Sokolov:2012p054105} We shall see that there is a class of cumulant parameterizations that avoid this fate, and our \ansatz\ is among them. Therefore, O$\lambda$DCT is guaranteed to be size-extensive, irrespective of how it is truncated.

Let us consider the size-extensivity of the DCT energy \eqref{eq:cumulant-energy}, but at first leave the cumulant parameterization abstract. The integrals $h$ and $\bar{g}$ are size-extensive, so it follows that the energy expression \eqref{eq:cumulant-energy} is also size-extensive, so long as the 1RDM and the 2RDM cumulant delivered by an approximate DCT method are both size-extensive. If the 2RDM cumulant is size-extensive, then its partial trace is size-extensive, so its eigenvalues must be extensive, so the 1RDM reconstructed by \eqref{eq:d-to-gamma-eval} must be as well.

The 2RDM cumulant, as defined by \eqref{eq:2rdm-decomp}, is not necessarily size-extensive when approximations are introduced. For example, computing the cumulant from \eqref{eq:2rdm-decomp} using RDMs delivered by truncated configuration interaction cannot be additively separable, for then the energy expression \eqref{eq:cumulant-energy} would imply the size-extensivity of truncated configuration interaction.

But let us suppose that the 2RDM cumulant is parameterized by some functional which is size-extensive, in agreement with the exact behavior. This still does not guarantee size-extensivity of the 2RDM cumulant. To compute the final 2RDM cumulant, we evaluate that functional \textit{at the parameters that satisfy the residual equations}. As pointed out previously,\cite{Szalay:1995p281} differentiating a connected energy functional need not give a connected residual, so the minimum energy parameters of separate computations of non-interacting subsystems need not be the minimum energy parameters of a calculation on all non-interacting subsystems at once. This spoils the size-extensivity of the theory.

Much as the variational formulation of the truncated UCC ansatz is not size-extensive for this reason,\cite{Szalay:1995p281} approximate OUDCT is not necessarily size-extensive for this reason. We have derived the residual expressions for the degree two doubles and quadruples OUDCT theory and identified a disconnected contribution to $r^{ijkl}_{abc}$, $P^{(ij/kl)}_{(ab/cd)} \bar{g}^{ij}_{ab} t^{kl}_{cd}$. We find a 2.5 mH size-consistency error for two infinitely separated Be atoms with the STO-3G basis set. The error is expected to be larger for larger basis sets and molecules.

Disconnected residual terms exist if there is a tensor contraction in the RDM expressions that becomes disconnected after differentiating a single amplitude. The resulting residual tensor contraction appears as the original tensor contraction, but with one amplitude to be removed. The requirement that all these terms to the energy residual be removed is exactly what motivated\cite{Szalay:1995p281} the formulation of SCVCC from VCC: SCVCC's amplitude residual consists of connected tensors only because the RDM formulas contain strongly connected tensors only.

Let us now assume that our DCT ansatz has amplitude residuals that consist of connected tensors only, i.e., the cumulant is parameterized by strongly connected tensor contractions. Then the cumulants are additively separable as desired. The orbital residuals are then guaranteed connected, as they are connected functions of the reduced density matrices. We can then rigorously say that an orbital-optimized DCT method is size-extensive when the cumulant parameterization consists only of strongly connected tensor contractions.

It is clear that using UCC or VCC as our cumulant parameterization will lead to size-extensivity error, and that SCVCC will not. What of the \ansatz\ ansatz? \textit{It too consists only of strongly connected tensor contractions, and will therefore give connected amplitude residuals and thus size-extensive energies.}

We know two proofs of this fact. The first relies on the general formula for weights in \ansatz, which is left to future research.\cite{Misiewicz2021:I} Here, we present the second. Recall that we derived \ansatz\ from the cumulant expressions of SCVCC. In our derivation, we derive the terms by repeatedly inserting a sum of strongly connected tensor contractions into strongly connected tensors. This can only produce more strongly connected tensors: if removing a single amplitude in the resulting diagrams would cause a disconnect, it would have caused a disconnect in the original diagrams as well. Each step of our iterative procedure yields strongly connected tensors, so the final cumualnt expression can only consist of strongly connected tensors, which is exactly what we needed to give size-extensive results. In this proof, \ansatz\ straightforwardly inherits its size-extensive residuals from the strongly connected cumulants of SCVCC theory.

\subsection{Explicit Equations}
\label{sec:explicit}

Now that we have chosen our exact ansatz, we may begin to approximate it. We only consider approximate theories where we include either all or none of the cumulant terms at a given degree of truncation in the amplitudes. Because the amplitudes are tensors of very large ranks, we cannot truncate up to $n$ amplitudes, but require separate degrees of truncation depending on the ranks of the tensors involved. As a crude estimate of which degrees of truncation to include, we appeal to M{\o}ller-Plesset perturbation theory (MPPT). A similar strategy was used in expectation value coupled cluster theory.\cite{Bartlett1988, Bartlett1989, Szalay:1995p281} We caution that MPPT is a heuristic designed for the limit of weak correlation and has relatively little to say about the tolerance to emergent strong correlation that ODC-12 possesses and we wish to maintain. Accordingly, we are unconcerned with a rigorous analysis within MPPT and assess MPPT's suggestions with the numerical benchmarks of Section \ref{sec:numerical}.

In our crude analysis, we shall assume that a rank $n$ cumulant element is of the same degree in the perturbation strength parameter $\mu$ as the rank $n$ coupled cluster amplitude.\footnote{mppt2}
We then collect all terms up to $\mathcal{O}(\mu^n)$.

We find that to get the energy terms of $\mathcal{O}(\mu^3)$, we require the rank two particle-hole cumulants as parameters at degree two, which recovers exactly the equations of the ODC-12 method. With this encouraging result, we continue on to recover the energy $\mathcal{O}(\mu^4)$. Let us begin with the contributions with doubles only. First, we need the cumulant with three doubles amplitudes. The OUDCT method adds several terms to the the $\lambda^{ij}_{ab}$ block of the cumulant,\cite{Sokolov:2014p074111} but O$\lambda$DCT adds no terms because we have $\lambda^{ij}_{ab} = \lambda^{ij}_{ab}$ to infinite order. This is another encouraging result, as the OUDCT ansatz deteriorates upon adding the degree three terms.\cite{Misiewicz:2020p244102} The O$\lambda$DCT ansatz does not suffer them.

Next, we need the 1RDM terms with four doubles amplitudes, which requires the cumulant to degree four in the doubles amplitudes.

\begin{align}
	\begin{split}
		\lambda^{ij}_{kl} =& - \frac{1}{2} \lambda^{ij}_{cd} \lambda^{cf}_{mn} \lambda^{mn}_{ef} \lambda^{de}_{kl} - \frac{1}{8} \lambda^{ij}_{cd} \lambda^{cd}_{mn} \lambda^{mn}_{ef} \lambda^{ef}_{kl} \\
		&- \frac{1}{8} P_-(ij) P_-(kl) [\lambda^{in}_{ef} \lambda^{ef}_{km} \lambda^{mj}_{cd} \lambda^{cd}_{nl}] \\
		&- \frac{1}{2} P_-(ij) P_-(kl) [\lambda^{im}_{cd} \lambda^{de}_{mk} \lambda^{jn}_{ef} \lambda^{fc}_{nl}]
	\end{split}\\[10pt]
	\begin{split}
		\lambda^{ab}_{cd} =& - \frac{1}{2} \lambda^{ab}_{kl} \lambda^{kn}_{ef} \lambda^{ef}_{mn} \lambda^{lm}_{cd} - \frac{1}{8} \lambda^{ab}_{kl} \lambda^{kl}_{ef} \lambda^{ef}_{mn} \lambda^{mn}_{cd} \\
		& - \frac{1}{8} P_-(ab) P_-(cd) [ \lambda^{af}_{mn} \lambda^{mn}_{ce} \lambda^{eb}_{kl} \lambda^{kl}_{fd} ] \\
		& - \frac{1}{2} P_-(ab) P_-(cd) [ \lambda^{ae}_{kl} \lambda^{lm}_{ec} \lambda^{bf}_{mn} \lambda^{nk}_{fd} ]
	\end{split} \\[10pt]
	\begin{split}
		\lambda^{ia}_{jb} =& - \frac{1}{2} \lambda^{im}_{bc} \lambda^{ce}_{kl} \lambda^{kl}_{de} \lambda^{ad}_{jm} - \frac{1}{2} \lambda^{ik}_{be} \lambda^{cd}_{km} \lambda^{lm}_{cd} \lambda^{ae}_{jl} \\ 
		& + \lambda^{ac}_{jk} \lambda^{kl}_{cd} \lambda^{de}_{lm} \lambda^{mi}_{eb} - \lambda^{ik}_{ec} \lambda^{ca}_{km} \lambda^{ed}_{jl} \lambda^{lm}_{db}\\
		& + \frac{1}{2} \lambda^{im}_{cd} \lambda^{de}_{mj} \lambda^{ca}_{kl} \lambda^{kl}_{eb} + \frac{1}{2} \lambda^{ae}_{kl} \lambda^{lm}_{eb} \lambda^{ki}_{cd} \lambda^{cd}_{mj}
	\end{split}
\end{align}

\noindent Before continuing with the perturbative analysis, comparing these new terms with the terms for the degree four OUDCT theory shows that the number of terms added to the energy functional has been reduced by about half. (These terms may be found in Appendix 1 of Reference \citenum{Sokolov:2014p074111}, although the reader is cautioned that the amplitudes written therein as $\lambda$ refer to unitary coupled cluster amplitudes, \textit{not} elements of the particle-hole cumulant.)

We now consider triples. First, a product of triples amplitudes generates terms of $\mathcal{O}(\mu^4)$ in the 1RDM, so we require the corresponding terms in the cumulant. Second, a triples amplitude and a doubles amplitudes can generate terms that are $\mathcal{O}(\mu^3)$ in the cumulant, for a $\mathcal{O}(\mu^4)$ contribution to the energy. Collectively, we need to add the terms that are degree two with doubles and triples. These terms, which follow, have computational scaling $\mathcal{O}(V^5 O^3)$ and can also be found in equations 31 and 32 of Reference \citenum{Sokolov:2014p074111}:

\begin{align}
	\begin{split}
		\lambda^{ij}_{kl} = \frac{1}{6} \lambda^{ijm}_{abc} \lambda^{abc}_{klm}
	\end{split}\\[10pt]
	\begin{split}
		\lambda^{ab}_{cd} = \frac{1}{6} \lambda^{abe}_{ijk} \lambda^{ijk}_{cde}
	\end{split}\\[10pt]
	\begin{split}
		\lambda^{ia}_{jb} = -\frac{1}{4} \lambda^{ikl}_{bcd} \lambda^{acd}_{jkl}
	\end{split} \\[10pt]
	\begin{split}
		\lambda^{ia}_{bc} = \frac{1}{2} \lambda^{ijk}_{bcd} \lambda^{ad}_{jk}
	\end{split} \\[10pt]
	\begin{split}
		\lambda^{ij}_{ka} = \frac{1}{2} \lambda^{ijl}_{abc} \lambda^{bc}_{kl}
	\end{split}
\end{align}

We lastly consider the $\mathcal{O}(\mu^5)$ energy correction. Within the O$\lambda$DCT ansatz, many RDM diagrams that could give rise to these terms are of particle-hole type and thus never arise in the underlying \ansatz\ ansatz. There is only one class of diagrams of this order that have not yet been included to generate $\mathcal{O}(\mu^4)$ 1RDM diagrams. These diagrams are contributions to the cumulant involving one triples amplitude and two doubles amplitudes. Including \textit{all} terms of three amplitudes, which may be doubles or triples, gives the following terms within O$\lambda$DCT:

\begin{align}
	\begin{split}
		\lambda^{ia}_{bc} = - \frac{1}{2} \lambda^{ade}_{jkl} \lambda^{jk}_{bd} \lambda^{il}_{ce} + \frac{1}{4} \lambda^{ade}_{jkl} \lambda^{jk}_{bc} \lambda^{il}_{de}
	\end{split} \\[10pt]
	\begin{split}
		\lambda^{ij}_{ka} = - \frac{1}{2} \lambda^{bcd}_{klm} \lambda^{il}_{bc} \lambda^{jm}_{ad} + \frac{1}{4} \lambda^{bcd}_{klm} \lambda^{ij}_{cd} \lambda^{lm}_{ab}
	\end{split}
\end{align}

This is an astounding result for two reasons.
\begin{enumerate}
	\item Normally, quadruples would be necessary to be correct to $\mathcal{O}(\mu^5)$ due to the contraction $\lambda^{ijkl}_{abcd} \lambda^{ab}_{ij} \bar{g}^{cd}_{kl}$. This term does not exist when the variables are particle-hole cumulants.
	\item In a unitary or variational coupled cluster theory, there are 14 distinct tensor contractions to the cumulant involving two doubles and one triple, or two triples and one double. By using \ansatz, we have reduced this number to 4, both of which have only one triples amplitude. This is much more economical.
\end{enumerate}

\section{Benchmark Computations}
\label{sec:numerical}

In the preceding section, we have shown that parameterizing the cumulant in terms of particle-hole cumulants rather than unitary coupled cluster amplitudes overcomes the formal objections to OUDCT. If the price for this is slow convergence of the energy with respect to the degree of ansatz truncation, O$\lambda$DCT would fare no better. 

This section reports data from pilot implementations of O$\lambda$DCT and related theories to assess the accuracy of O$\lambda$DCT for manageable ansatz truncations. In Section \ref{sec:nomenclature}, we state our procedure and our naming conventions for the methods we benchmark. The remaining sections are numerical studies.

\subsection{Procedure and Nomenclature}
\label{sec:nomenclature}

One of our objectives is to assess how competitive O$\lambda$DCT methods are among Hermitian orbital-optimized methods that variationally minimize a parameterization of the 2RDM and have a triples correction. This requires defining many methods.

For UCC, VCC, SCVCC, and \ansatz, we call the corresponding energy methods \textit{not using DCT} OVUCC, OVCC, OSCVCC, and O$\lambda$, respectively. For the methods \textit{using DCT}, we call them OUDCT, OVDCT, OSVDCT, and O$\lambda$DCT.

For the methods not using DCT, we shall study only those methods where the RDMs are truncated up to some degree in the amplitudes. For the methods using DCT, we shall study only methods where the 2RDM cumulant is truncated up to a particular degree in the amplitudes, and the same cumulant truncation is used for the matrix $d$ in the 1RDM reconstruction, \eqref{eq:d-to-gamma-eval}. A DCT method must agree with its non-DCT counterpart in the untruncated limit, neglecting questions of convergence.

We end the name of a method with $n$, i.e., OVCC$n$, to indicate it contains all doubles terms of the ansatz up to degree $n$. To signify that a method includes all doubles and triples terms up to degree $n$ in the amplitudes, we add ``+T$n$" to the method-name. For example, adding a degree two doubles correction to the unitary theory with DCT and degree four doubles yields OUDCT4+T2.

Although different ans{\"a}tze in general have different terms, these terms are sometimes coincident. In the special case of doubles at degree two and no higher rank tensors, all four DCT theories predict the same equations, which we call ODC-12, and all four Taylor series theories predict the same equations, which are those of orbital-optimized linearized coupled cluster doubles (OLCCD).\cite{Bozkaya:2013p054104} We shall use these names when we do not need to associate the methods with one particular ansatz. Furthermore, all four DCT theories predict the same degree two triples terms. There are also coincidences among VCC and SC-VCC theories: OVCC$n$ and OSCVCC$n$ are identical, OVDCT$n$ and OSVDCT$n$ are identical, and OVDCTn+T2 and OSVDCTn+T2 are identical.

With our methods defined, we now outline how we performed the benchmarks. We created a code generator to draw all fully connected Brandow diagrams. The diagram then computed the tensor contractions for elements of the reduced density matrices and attached each theory's specific prefactor, using Nauty\cite{mckay2013practical} to compute the automorphism groups needed for VCC prefactors. An energy expression may be derived from this, which is differentiated to construct the amplitude residual equations.

The automatically generated equations for the residuals were computed using the \texttt{opt\_einsum} package\cite{G_A_Smith:2018p753} for tensor contractions and \textsc{Psi4} 1.4 for all integrals.\cite{Smith:2020p184108, Smith:2018p3504} The residual steps were computed using the standard first-order update in terms of orbital energy denominators for Taylor series first, and using elements of $\tilde{F}$ for DCT theories.\cite{Wang:2016p4833} Convergence was accelerated by direct inversion of the iterative subspace (DIIS).\cite{Hamilton:1986p5728} We required that the root mean square gradient for all geometry optimizations was $1 \times 10^{-6}$ or less, and all energy single points had amplitude gradient under $1 \times 10^{-6}$, by which point the energy was converged to within $1 \times 10^{-10}$. When comparing tensors defined in separate orbital spaces, we rotate one tensor into the orbital space of the other and compute a Euclidean norm, as described previously.\cite{Misiewicz:2020p244102}

As a correctness check on our results, we computed the dipole both by finite difference of energies and from our analytic RDMs. In all cases, the results were identical within our numerical tolerance. The correctness check on the unitary coupled cluster equations has been described previously.\cite{Misiewicz:2020p244102} We established the correctness of the variational coupled cluster energy expression by using the \textsc{Forte} package\cite{Forte2020} of Evangelista and coworkers to compute variational coupled cluster using cluster operators, rather than tensor contractions, up to degree five. Our energies matched to ten decimal places. We confirmed the strongly-connected variational coupled cluster expressions by confirming that the terms marked for inclusion in variational coupled cluster but exclusion from variational coupled cluster were indeed weakly connected. For the \ansatz\ ansatz, we repeat the derivations ourselves in Section 1 of the Supporting Information.

\subsection{\ce{H2 Dissociation}}
\label{ssec:h2}

We begin with a study of \ce{H2}. This is a two-electron system, so triples contributions are non-existent. While our formal analysis in Section \ref{sec:formal-h2} proved that O$\lambda$DCT is exact at degree two in a minimal basis set, we want to see if the good performance continues in a larger one-particle basis set.

\begin{figure}
	\caption{Error in the dissociation curves of \ce{H2} from 0.6 \AA\ to 3.5 \AA\ computed with low-degree truncations of the O$\lambda$DCT ansatz in the cc-pVDZ basis set. Errors are relative to full configuration interaction.}
	\includegraphics[width=0.8\linewidth]{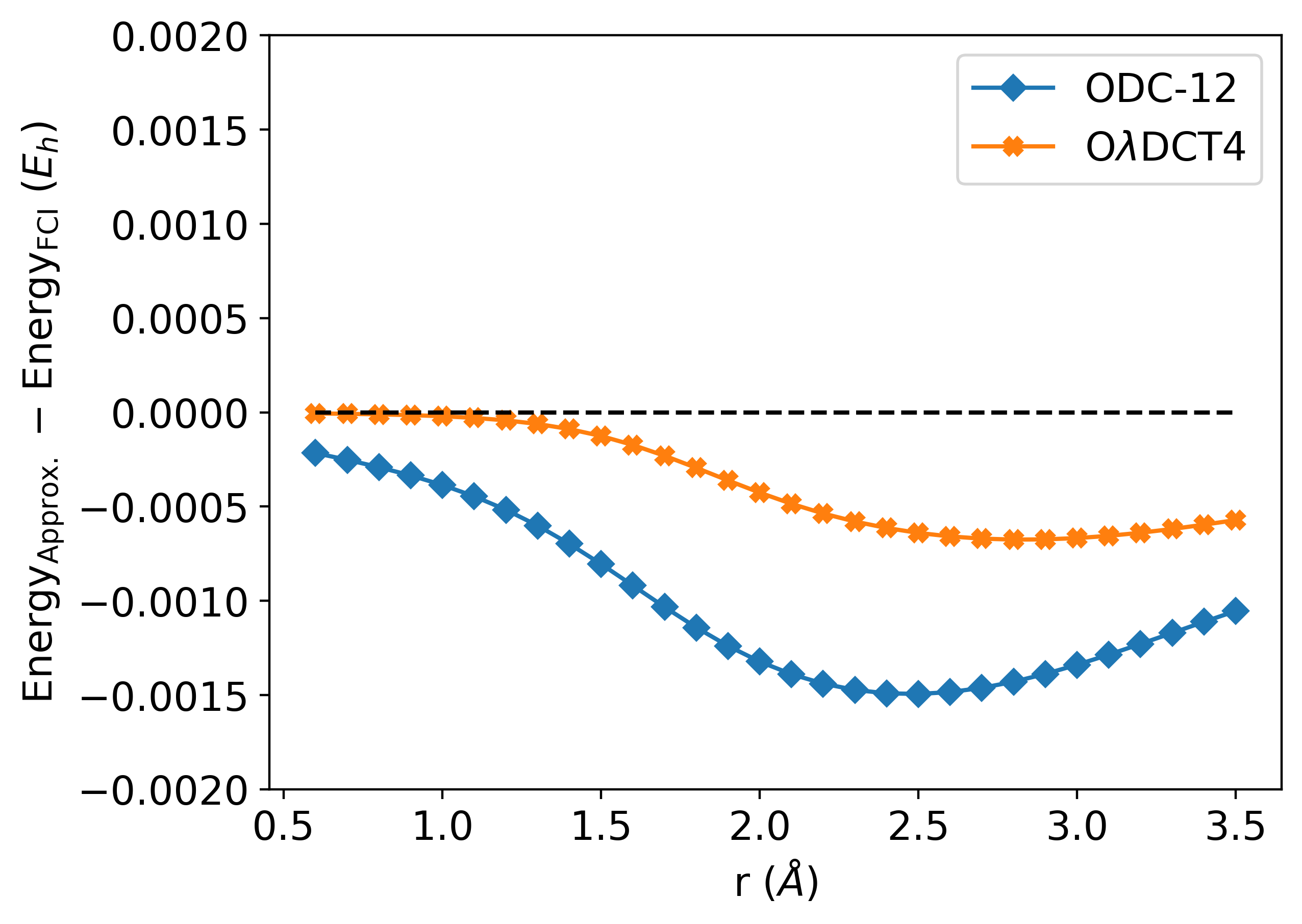}
	\label{fig:h2-energy}
\end{figure}

We first examine whether O$\lambda$DCT4 for \ce{H2} improves the accuracy of the O$\lambda$DCT2 (ODC-12) dissociation curve, using the cc-pVDZ basis set.\cite{Dunning:1989p1007} The errors in the dissociation curves are shown in Figure \ref{fig:h2-energy}.

We observe that the O$\lambda$DCT4 energy curve is more accurate \textit{across the entire region considered}. At worst, the ODC-12 energy curve is 1.49 mH below FCI, while O$\lambda$DCT4 has an error of at worst 0.67 mH, reducing the error to less than half. This contrasts starkly with the counterpart OUDCT4 approximation for the same problem, which has more severe errors and eventually diverges.\cite{Misiewicz:2020p244102} We also observe that O$\lambda$DCT4 is more accurate than ODC-12 in the weakly correlated limit. At 0.6 \AA, ODC-12 is in error by 216 $\mu$H, compared to the 6 $\mu$H error of O$\lambda$DCT4. This supports the hypothesis that up to the order of truncation, our cumulant parameterization corresponds to \textit{some} $n$-representable parameterization.

\begin{figure}
	\caption{The difference between the converged doubles amplitudes for O$\lambda$DCT theories and the exact off-diagonal cumulants as a fraction of the norm of the exact off-diagonal cumulants for \ce{H2} computed with the cc-pVDZ basis set. All occupied and virtual orbitals are natural spinorbitals.}
	\includegraphics[width=0.8\linewidth]{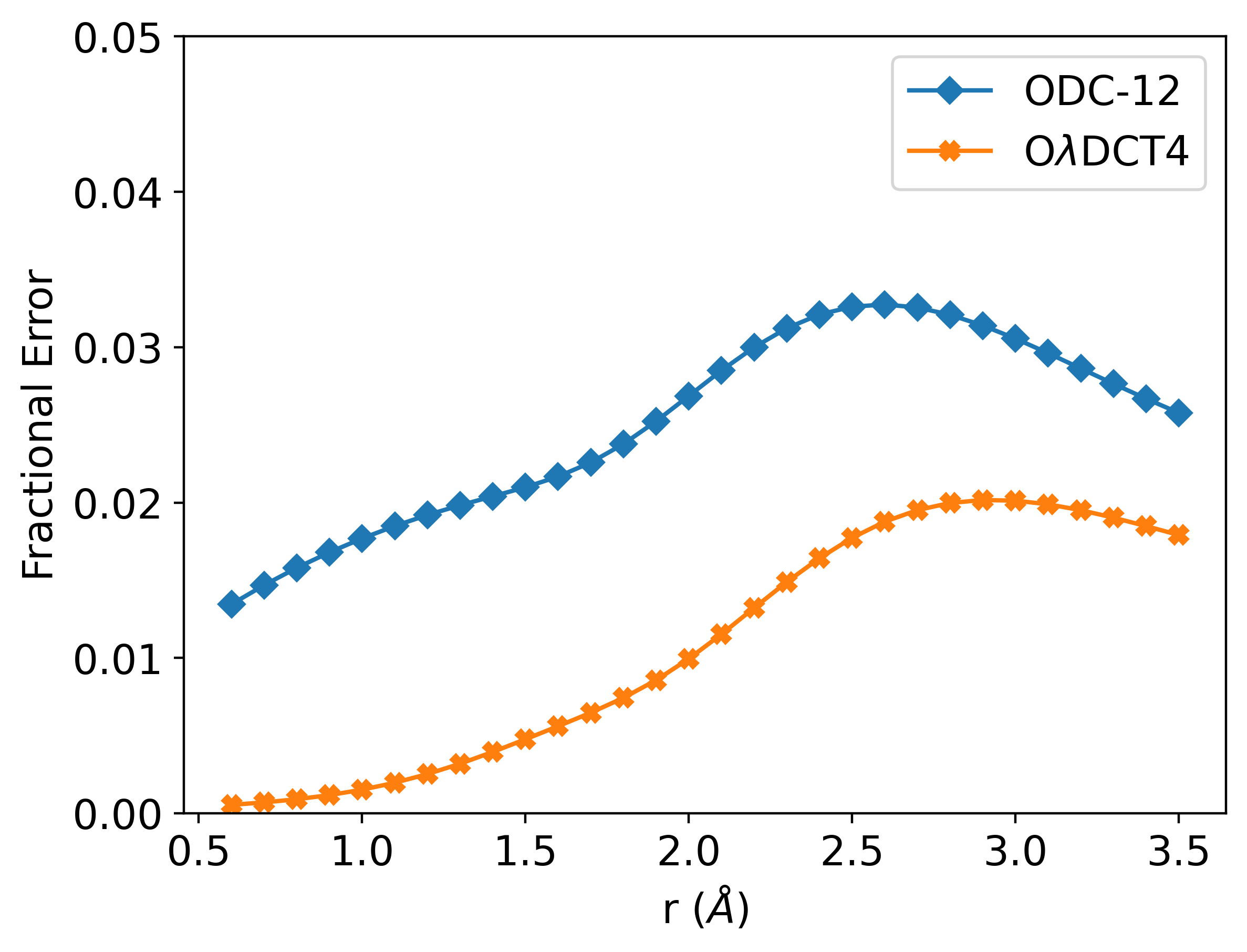}
	\label{fig:h2-error}
\end{figure}

To confirm that the variables of \ansatz\ are cumulants, we consider the difference between the converged amplitudes of O$\lambda$DCT4 and the theoretical exact parameters of the orbital-optimized form of \ansatz, i.e., the off-diagonal blocks of the cumulant in the basis of natural spinorbitals. We have performed the same analysis for ODC-12. This is shown in Figure \ref{fig:h2-error}.

For both theories, the fractional error in the amplitudes remains small. It reaches a maximum of 3.2\% for ODC-12 and 2.0\% for O$\lambda$DCT4. This suggests that, as expected, including more terms from the O$\lambda$DCT ansatz drives the amplitudes towards those of the exact ansatz. This effect is especially strong in the weakly correlated limit, as expected. It also suggests that the success of ODC-12 for \ce{H2} should be attributed to its variables mirroring the off-diagonal cumulant of the exact O$\lambda$DCT ansatz and not, for example, the unitary coupled cluster coupled amplitudes of the exact OUDCT ansatz. In comparison, the error in the ODC-12 amplitudes \textit{relative to the amplitudes of the exact OUDCT ansatz} grows along the dissociation curve, reaching even 33.7\% by 3.5 \AA. This data is available in Section 2 of the Supporting Information.

\begin{figure}
	\caption{Error in the dissociation curves of \ce{H2} from 0.6 \AA\ to 2.5 \AA\ in the cc-pVDZ basis set, computed with degree four truncations of the orbital-optimized unitary coupled cluster, variational coupled cluster, and \ansatz\ ans{\"atze}, with and without density cumulant theory. Errors are relative to full configuration interaction.}
	\includegraphics[width=0.8\linewidth]{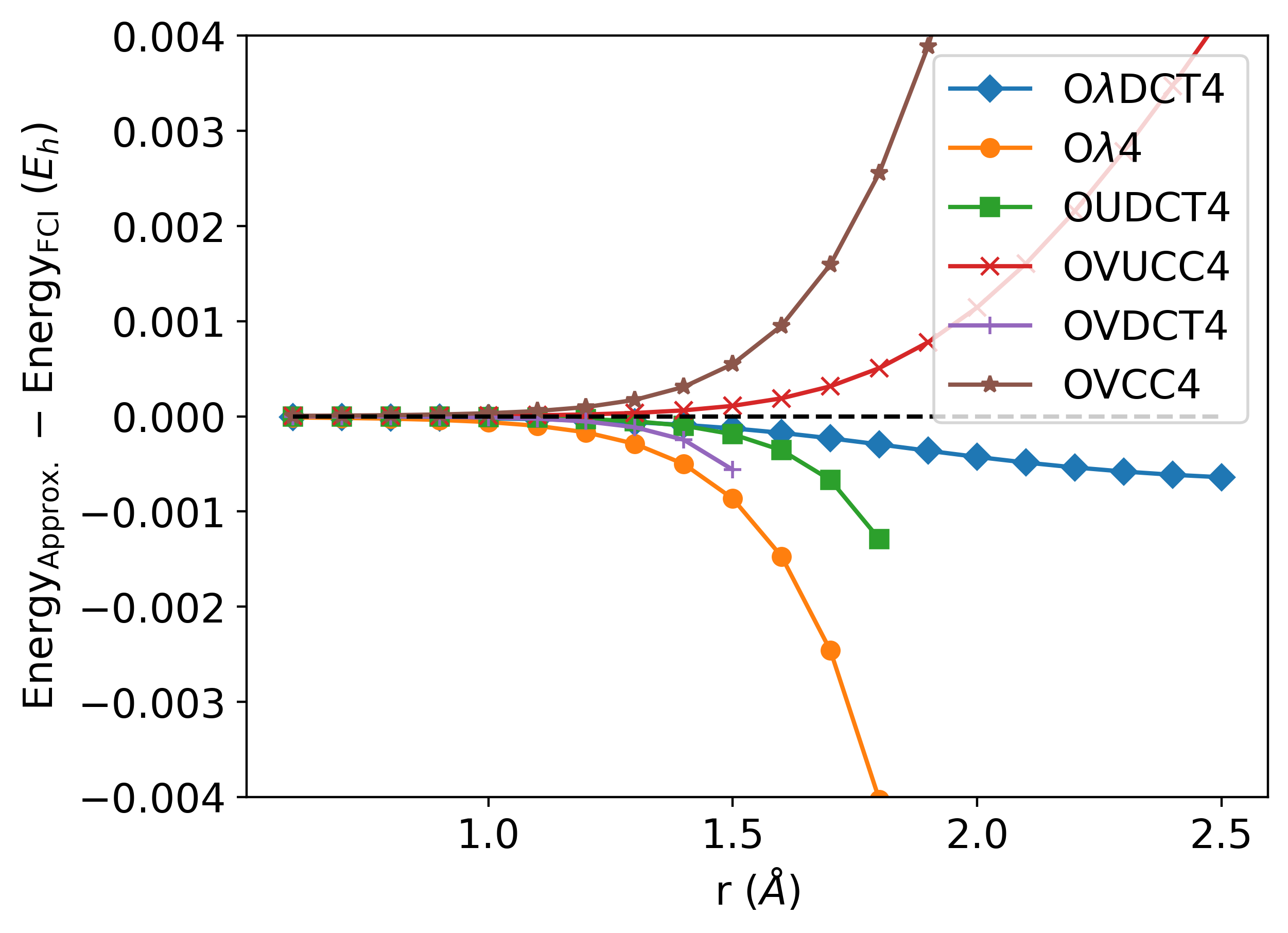}
	\label{fig:oo-h2-en}
\end{figure}

Having thoroughly shown the relation between our O$\lambda$DCT4 and the cumulants, we now turn to look at the competitiveness of O$\lambda$DCT4. We compare the accuracy of this curve to the accuracy of orbital-optimized theories starting from the unitary, variational, and \ansatz\ ans{\"a}tze at degree four in doubles, using both DCT and the more straightforward Taylor series truncation. Our results are illustrated by Figure \ref{fig:oo-h2-en}.

Using O$\lambda$DCT produces a curve an order of magnitude better than any other, far from equilibrium. Using only DCT with some other ansatz or using \ansatz\ without DCT tends to produce a rapid decrease in the energy shortly before a divergence somewhere between 1.6 \AA\ and 2 \AA. We interpret this as \ansatz\ and DCT balancing each other. Using either variational or unitary coupled cluster theory with the traditional Taylor series truncation avoids divergence, but the quality of the energy deteriorates markedly along the curve, with variational coupled cluster deteriorating the most. The relevant data is provided in Section 2 of the Supporting Information.

We close this subsection with a technical note. Previous DCT studies of the \ce{H2} dissociation curve stop before 2.5 \AA,\cite{Copan:2018p4097, Sokolov:2014p074111, Misiewicz:2020p244102} while our study goes to 3.5 \AA. Converging these values required damping the DCT update steps. Not doing so led to large cumulant updates which in turn led to the cumulant partial trace, $d$, having eigenvalues less than $-0.25$. Consequently, reconstruction of the 1RDM by \eqref{eq:d-to-gamma-eval} failed. The large cumulant updates originate in the denominators of the amplitude gradient, \eqref{eq:amplitude-gradient}. As discussed in Section \ref{subsec:block-diag}, when natural orbital occupation numbers approach 0.5, there will be small denominators. This leads to large cumulant changes. Alternate convergence algorithms are recommended for future DCT studies on systems with occupation numbers far from the ideal 0 and 1.

\subsection{Size-Extensivity Errors}

The remaining benchmarks include iterative triples corrections. As discussed in Section \ref{subsec:oudct-size-extensivity}, not all these theories have connected residual gradients, and this introduces the possibility of size-extensivity errors. We see disconnected residual terms in the OVUCCn+T2, OVCCn+T2, OVDCTn+T3, and OVUCCn+T3 theories.  These terms are present in the exact theories and cannot be removed. We reiterate that O$\lambda$DCT, our primary theory of interest, contains no such terms.

Before benchmarking such methods on actual systems, we wish to benchmark the severity of the size-extensivity errors. Because the doubles-only theories have connected terms only, we expect the errors to be most acute for systems with highly signficant triples. We first attempted to benchmark this effect using two infinitely separated \ce{CN} radicals, but the equations for a single \ce{CN} proved difficult to converge, and we could not surmount the convergence problems with two \ce{CN} radicals. We therefore switched to a high-spin system of two \ce{NF} radicals. For two \ce{NF} radicals at a bond length of 1.33 \AA, we report the following size-extensivity errors in the cc-pVDZ basis set.

\begin{table}[h]
	\begin{tabular}{c | c}
		Method & Error ($\mu E_h$)  \\ \hline
		OVUCC3+T2 & $7 \times 10^{1}$ \\
		OVCC3+T2 & $3 \times 10^{2}$ \\
		OVUCC4+T2 & $9 \times 10^{1}$ \\
		OVCC4+T2 & $5 \times 10^{2}$ \\
		OVUCC3+T3 & $7 \times 10^{-1}$ \\
		OVCC3+T3 & $2 \times 10^{1}$ \\
		OVUCC4+T3 & $7 \times 10^{-1}$ \\
		OVCC4+T3 & $3 \times 10^{1}$ \\
		OUDCT3+T3 & $6 \times 10^{1}$ \\
		OVDCT3+T3 & $2 \times 10^{2}$ \\
		OUDCT4+T3 & $8 \times 10^{1}$ \\
		OVDCT4+T3 & $3 \times 10^{2}$ \\
		
	\end{tabular}
\caption{Size-extensivity errors for two infinitely separated \ce{NF} radicals with \ce{N-F} bond length 1.33 \AA, computed with the cc-pVDZ basis set.}
\end{table}

The errors we observe for these simple systems are relatively mild, at under a millihartree. We observe that the errors are consistently larger for variational methods than for unitary methods. This is a trend across all of our numerical results and can be predicted\cite{Kutzelnigg_1977, Kutzelnigg:1991p349} from the reduced density matrix expressions, as terms in the unitary coupled cluster theory always have a smaller weight than in the variational coupled cluster theory.  We also observe that the effect decreases as more triples terms are included. This is unsurprising, as the exact theory must be size-extensive. We also see that the use of DCT \textit{increases} the magnitude of the size-extensivity error. While the mechanism for this is not clear to us, it discourages using variational coupled cluster or unitary coupled cluster with DCT.

\subsection{\ce{Be2} Dissociation}

We next consider dissociation of the beryllium dimer with the cc-pVDZ basis set.\cite{Dunning:1989p1007} This system is small enough that reference data from full configuration interaction can be computed. The beryllium dimer is also known to be highly sensitive to the electron correlation treatment used,\cite{Mullinax_2015, El_Khatib:2014p6664, Ascik:2011p074110, Sharma:2014p104112, Sosa:1988p5974, Lee:1984p4371, Shirley:1991p588, Bartlett:1990p513} and so shall motivate our inclusion of triples terms.

Full configuration interaction reports a minimum energy geometry at 2.56 \AA, which corresponds to the experimental minimum at 2.45 \AA.\cite{Merritt:2009p1548} ODC-12, which is the simplest O$\lambda$DCT description, agrees qualitatively, placing the minimum at 2.61 \AA. Any improvement on ODC-12 must retain this qualitative feature.

For the sake of economy, let us first use the theories with only doubles amplitudes. Not one of our DCT parameterizations (OUDCT, OVDCT, O$\lambda$DCT) is able to reproduce a minimum by inclusion of degree three or four doubles. We can attempt the same with their Taylor series counterparts (OVUCC, OVCC, O$\lambda$) and find that most of them also lose the minimum. While O$\lambda$4 still has a minimum, its predicted bond length is \textit{less} accurate than ODC-12's.

To improve the description of the minimum, each ansatz requires more terms. In this study, we add the next non-zero degree of triples rather than add more doubles. We have two reasons to favor triples, in addition to the perturbation theory arguments of Section \ref{sec:explicit}:
\begin{enumerate}
	\item It has been previously observed that even CCSD, where the only approximation is truncating the cluster operator, cannot correctly predict the existence of a bound \ce{Be2} structure around 2.5 \AA.\cite{Mullinax_2015, Bartlett:1990p513, Shirley:1991p588, Lee:1984p4371, Sosa:1988p5974, Ascik:2011p074110} For CCSD, only one remedy is possible: use higher-order excitations. And indeed, methods that account for connected triples\cite{Mullinax_2015, Bartlett:1990p513, Shirley:1991p588, Lee:1984p4371, Sosa:1988p5974, Sharma:2014p104112, Ascik:2011p074110} correctly predict a bound structure.
	\item Our previous study of the unitary ansatz with and without DCT\cite{Misiewicz:2020p244102} concluded that inclusion of triples amplitudes were needed to improve the accuracy for equilibrium properties beyond that of ODC-12, and that addition of further doubles terms without higher excitations would decrease accuracy.
\end{enumerate}

\begin{table}
	\caption{ Errors in the equilibrium bond length and harmonic vibrational frequency of \ce{Be2}, relative to FCI, for approximate orbital-optimized variational methods, using the cc-pVDZ basis set. n.c. means that we could not converge to a minimum. }
	\label{tbl:be2-eq}
	\centering
	\begin{subtable}[t]{.8\linewidth}%
		\caption{$\Delta r_e$ (pm), without triples}
		\centering%
		\begin{tabular}{c|cccccccc}
			Degree & OVUCC & OVCC & OSCVCC & O$\lambda$ & OUDCT & OVDCT & OSVDCT & O$\lambda$DCT \\ \hline
			2 & n.c. & n.c. & n.c. & n.c. & 5.86 & 5.86 & 5.86 & 5.86 \\
			3 & n.c. & n.c. & n.c. & n.c. & n.c. & n.c. & n.c. & N/A \\
			4 & n.c. & n.c. & n.c. & 12.94 & n.c. & n.c. & n.c. & n.c. \\
		\end{tabular}
	\end{subtable} \\
	\begin{subtable}[t]{.8\linewidth}
		\centering
		\caption{$\Delta r_e$ (pm), with T2 iterative triples}
		\begin{tabular}{c|cccccccc}
			Degree & OVUCC & OVCC & OSCVCC & O$\lambda$ & OUDCT & OVDCT & OSVDCT & O$\lambda$DCT \\ \hline
			2 & n.c. & n.c. & n.c. & n.c. & --9.01 & --9.01 & --9.01 & --9.01 \\
			3 & 11.35 & n.c. & n.c. & n.c. & n.c. & n.c. & n.c. & N/A \\
			4 & --1.01 & 4.48 & 7.13 & --13.39 & --2.83 & n.c. & n.c. & --0.32 \\
		\end{tabular}
	\end{subtable}\\
	\begin{subtable}[t]{.8\linewidth}
		\centering
		\caption{$\Delta r_e$ (pm), with T3 iterative triples}
		\begin{tabular}{c|cccccccc}
			Degree & OVUCC & OVCC & OSCVCC & O$\lambda$ & OUDCT & OVDCT & OSVDCT & O$\lambda$DCT \\ \hline
			3 & 17.09 & n.c. & n.c. & n.c. & n.c. & n.c. & n.c. & --9.04 \\
			4 & 2.44 & 14.05 & 13.63 & --14.05 & --0.17 & n.c. & n.c. & --0.81 \\
		\end{tabular}
	\end{subtable}
	\begin{subtable}[t]{.8\linewidth}%
		\caption{$\Delta \omega$ (\cm), without triples}
		\centering%
		\begin{tabular}{c|cccccccc}
			Degree & OVUCC & OVCC & OSCVCC & O$\lambda$ & OUDCT & OVDCT & OSVDCT & O$\lambda$DCT \\ \hline
			2 & n.c. & n.c. & n.c. & n.c. & --34 & --34 & --34 & --34 \\
			3 & n.c. & n.c. & n.c. & n.c. & n.c. & n.c. & n.c. & N/A \\
			4 & n.c. & n.c. & n.c. & --107 & n.c. & n.c. & n.c. & n.c. \\
		\end{tabular}
	\end{subtable} \\
	\begin{subtable}[t]{.8\linewidth}
		\centering
		\caption{$\Delta \omega$ (\cm), with T2 iterative triples}
		\begin{tabular}{c|cccccccc}
			Degree & OVUCC & OVCC & OSCVCC & O$\lambda$ & OUDCT & OVDCT & OSVDCT & O$\lambda$DCT \\ \hline
			2 & n.c. & n.c. & n.c. & n.c. & 76 & 76 & 76 & 76 \\
			3 & --73 & n.c. & n.c. & n.c. & n.c. & n.c. & n.c. & N/A \\
			4 & 3 & --31 & --45 & 76 & 7 & n.c. & n.c. & 1 \\
		\end{tabular}
	\end{subtable}\\
	\begin{subtable}[t]{.8\linewidth}
		\centering
		\caption{$\Delta \omega$ (\cm), with T3 iterative triples}
		\begin{tabular}{c|cccccccc}
			Degree & OVUCC & OVCC & OSCVCC & O$\lambda$ & OUDCT & OVDCT & OSVDCT & O$\lambda$DCT \\ \hline
			3 & --103 & n.c. & n.c. & n.c. & n.c. & n.c. & n.c. & 80 \\
			4 & --16 & --84 &--82 & 85 & --7 & n.c. & n.c. & 7 \\
		\end{tabular}
	\end{subtable}
\end{table}

We now add the degree two and degree three triples terms to all theories and attempt to compute the equilibrium geometries and harmonic frequencies. The errors for all combinations are shown in Table \ref{tbl:be2-eq}. We see that even with triples, many theories fail to predict a minimum. Any attempt to add triples to OLCCD fails, and we see many failures if degree four doubles are not included. Degree three triples do not change whether the theory reports a minimum. These findings are all expected from our earlier arguments from M{\o}ller-Plesset perturbation theory. While OVUCC3+T2 and OVUCC3+T3 find a qualitatively correct minimum, they have relatively high errors and are inferior to the much simpler ODC-12.

Turning to the theories with degree four doubles, we find that most theories at least report a minimum, except for OVDCT and OSVDCT. In these cases, we were not able to converge the equations anywhere near equilibrium. We suspect problems of the kind seen for \ce{H2} with OUDCT.\cite{Misiewicz:2020p244102} Of those theories that do converge, we observe that including degree three triples terms tends to \textit{worsen} the equilibrium properties. For example, O$\lambda$DCT4+T2 degrades from --0.32 pm bond distance error to --0.81 pm and from 1 \cm\ harmonic vibrational frequency error to 7 \cm\ error. But this is a relatively minor degradation in comparison to the 5.86 pm and --34 \cm\ errors of ODC-12. The +T3 terms improve one theory: when OUDCT4+T2 becomes OUDCT4+T3, the frequency error flips sign, and the bond distance error diminishes from --2.83 pm to --0.17 pm. We refrain from judging the worth of the +T3 terms until after the next subsection.

\begin{figure}
	\caption{Non-parallelism error (NPE) of the \ce{Be2}/cc-pVDZ dissociation curve, using infinite separation as the zero point.}
	\includegraphics[width=0.9\linewidth]{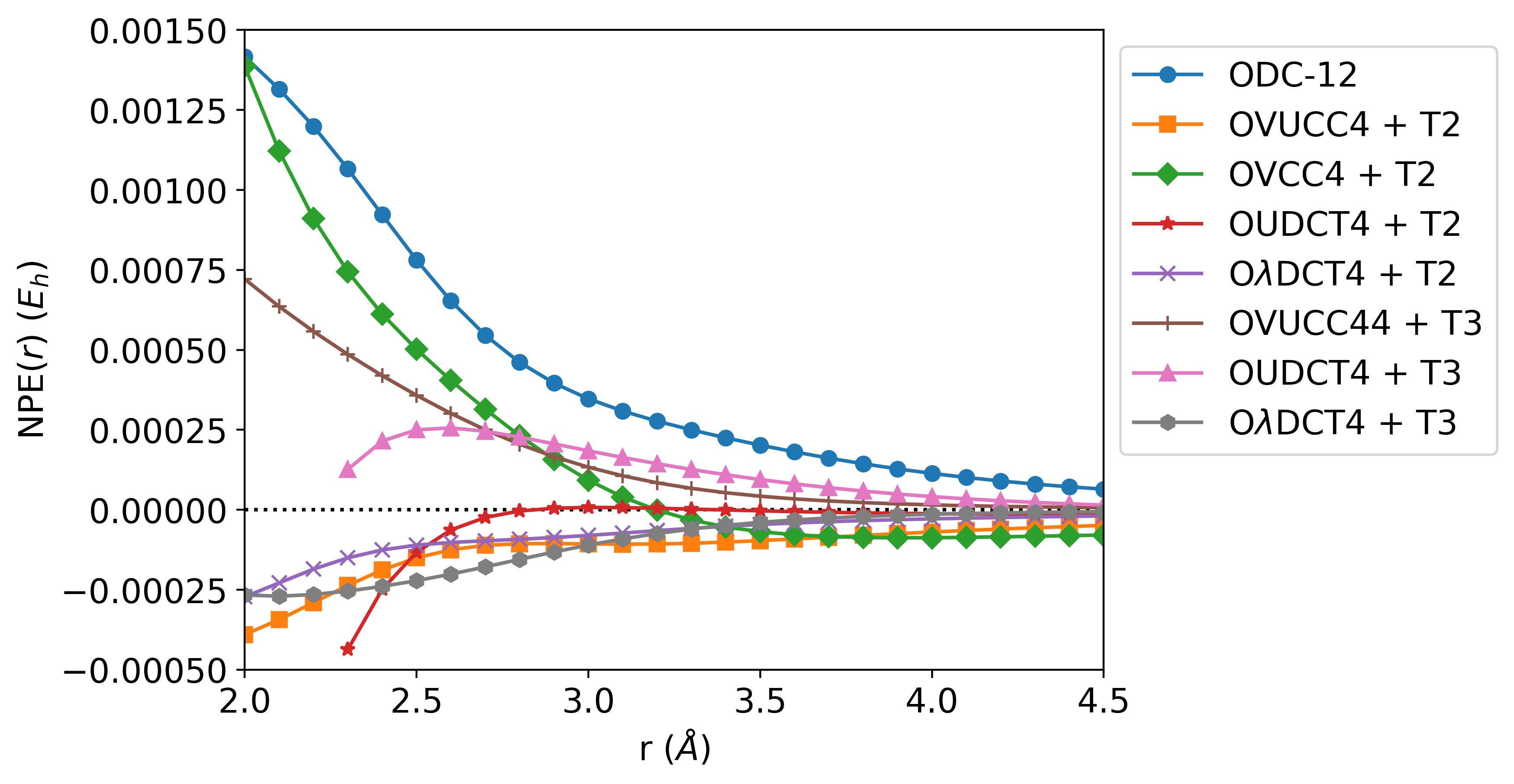}
	\label{fig:be2-npe}
\end{figure}

We now wish to see how the most accurate theories describe the dissociation curves, to ascertain which improve upon ODC-12 for both description of the entire curve, in addition to equilibrium properties. In Figure \ref{fig:be2-npe}, we compare the non-parallelism error of the theories which are at least as accurate for ODC-12 for either property. The non-parallelism error is defined as $[E\textsubscript{Approx.} - E\textsubscript{FCI}](r) - [E\textsubscript{Approx.} - E\textsubscript{FCI}](r \to \infty)$. The curves for all qualitatively accurate theories are given in Section 3 of the Supporting Information.

For all seven of these methods, \textit{when they converge}, they have smaller non-parallelism errors than ODC-12. The OUDCT based theories could not be converged for bond lengths smaller than 2.3 \AA, which we suspect to be the same problem observed in \ce{H2}.\cite{Misiewicz:2020p244102} For a small non-parallelism error and consistent convergence, the three best theories are O$\lambda$DCT4+T2, O$\lambda$DCT4+T3, and OVUCC4+T2. Of the three best theories, two are of the O$\lambda$DCT ansatz. O$\lambda$DCT4+T2 is more accurate than OVUCC4+T2 across the entire curve. O$\lambda$DCT4+T2 and O$\lambda$DCT4+T3 have a tradeoff in how the error grows. Error in O$\lambda$DCT4+T2 grows for short bond distances, with no sign of slowing down. Error in O$\lambda$DCT4+T3 has the same character as observed in O$\lambda$DCT4 for \ce{H2}, with increase in error being strongest in the middle, before flattening out in the region expected to be most multiconfigurational.

\begin{figure}
	\caption{The difference between the converged doubles amplitudes for O$\lambda$DCT theories and the exact off-diagonal cumulants as a fraction of the norm of the exact off-diagonal cumulants for \ce{Be2} computed with the cc-pVDZ basis set. All occupied and virtual orbitals are natural spinorbitals. ODC-12+T2 and ODC-12+T3 errors are too similar to be distinguished on this graph.}
	\includegraphics[width=0.8\linewidth]{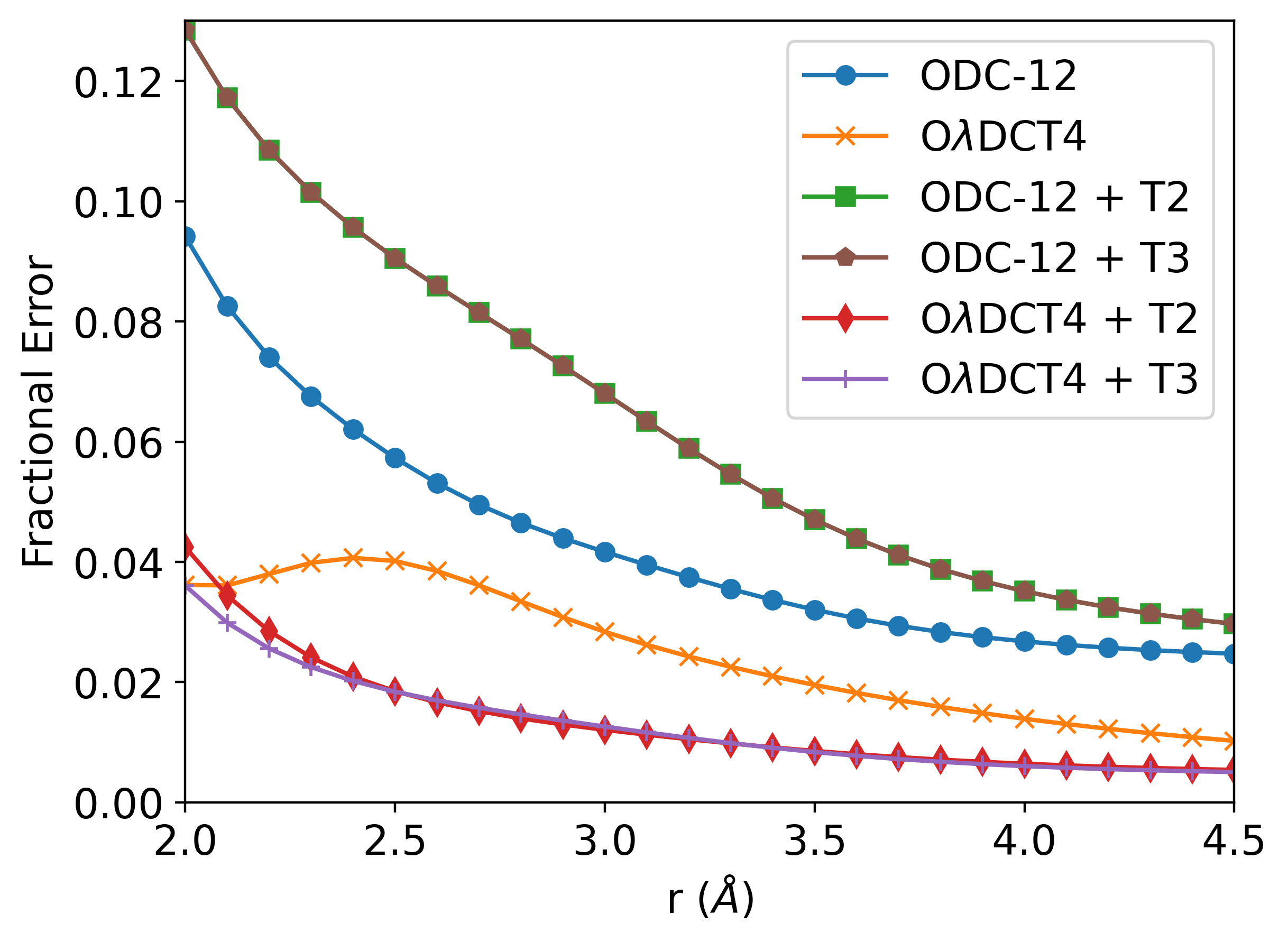}
	\label{fig:be2-lambda}
\end{figure}

Lastly, to confirm that the accuracy of the O$\lambda$DCT methods is correlated with their variables approximating cumulants, we plot the error in the amplitudes compared to the error of the exact O$\lambda$DCT ansatz in Figure \ref{fig:be2-lambda}. Numerical values are reported in Section 3 of the Supporting Information.

The addition of the degree four doubles terms of O$\lambda$DCT uniformly decreases the amplitude errors compared to ODC-12. In the highly compressed region, further compression means the O$\lambda$DCT4 amplitude error decreases. We are not aware of an interpretation of this fact, but point out that the energy error in this region of the O$\lambda$DCT4 surface is quite large. This is more evidence that our ansatz drives the variables to be cumulants, but also shows that this does not necessarily lead to an increase in accuracy.

Adding the T2 terms decreases the error in the doubles amplitudes \textit{even further} for most of the surface, excluding the aforementioned highly compressed region. Adding the T3 terms has only a minor effect on how well the doubles amplitudes approximate the doubles amplitudes of the exact theory, but they decrease the error when added to O$\lambda$DCT4+T2, to below that of O$\lambda$DCT4.

\subsection{Equilibrium Properties of Diatomics}

We now study energies of diatomics using our approximate theories, using CCSDTQ(P) energetics as reference values. Compared to our study of OUDCT, we have removed \ce{BeO} from our test set and replaced it with \ce{BF}, as the equilibium properties of \ce{BeO} had a large difference compared to CCSDTQP. We also report data for \ce{F2}, although we exclude it from all statements about average data.

Our analysis groups theories together, depending on how they treat triples.

\subsubsection{Doubles Only}

In a previous study,\cite{Misiewicz:2020p244102} we showed that including more OUDCT doubles terms tended to \textit{worsen} equilibrium properties of diatomics compared to ODC-12 or orbital-optimized linearized coupled cluster doubles (OLCCD). To assess \ansatz\ as a new DCT ansatz, our first question is to assess whether this conclusion holds for O$\lambda$DCT terms. We find that it does, as it does for the other ans{\"a}tze, whether truncated as Taylor series or whether their truncated cumulants are fed into DCT. For example, OLCCD has mean average errors of 0.35 pm and 28 \cm\ for equilibrium bond lengths and harmonic vibrational frequencies, respectively. ODC-12 has 0.64 pm and 50 \cm, while O$\lambda$DCT4 has 0.70 pm and 56 \cm. Adding more doubles terms always leads to worse results, as we show in Section 4 of the Supporting Information.

To develop parameterizations more accurate than ODC-12, it appears necessary to include triples.

\subsubsection{Degree Two Triples}

We nexr benchmark methods with the +T2 iterative triples terms. We find that O$\lambda$DCT4+T2 has a mean absolute error of 0.27 pm for equilibrium bond lengths and 27 \cm\ for harmonic vibrational frequencies, for our eight diatomic test set. This roughly halves the error from ODC-12.

To analyze precisely how accurate O$\lambda$DCT is relative to other orbital-optimized theories that approximate an exact parameterization of the reduced density matrices, we report errors in the geometries in Table \ref{tbl:diatomic_geom_t2} and errors in the harmonic vibrational frequencies in Table \ref{tbl:diatomic_freq_t2}. All methods considered use their degree two truncation of triples.

\begin{table}
	\singlespacing
	\centering
	\begin{tabular}{cc|cccccccc}
		& & OVUCC & OVCC & OSCVCC & O$\lambda$ & OUDCT & OVDCT & OSVDCT & O$\lambda$DCT \\ \hline
		\multirow{3}{*}{\ce{N2} $\begin{dcases*} \\ \\ \end{dcases*}$} & 2 & 0.95 & 1.06 & 0.89 & 0.89 & 0.34 & 0.34 & 0.34 & 0.34\\
		& 3 & --0.16 & --0.34 & --0.43 & N.A.        & --0.39 & --0.58 & --0.58 & N.A.\\
		& 4 & 0.36 & 0.46 & 0.31 & 0.33 & 0.27 & 0.31 & 0.31 & 0.26\\ \hline
		\multirow{3}{*}{\ce{CO} $\begin{dcases*} \\ \\ \end{dcases*}$} & 2 & 0.75 & 0.92 & 0.66 & 0.66 & 0.28 & 0.28 & 0.28 & 0.28\\
		& 3 & 0.04 & --0.05 & --0.20 & N.A.        & --0.22 & --0.36 & --0.36 & N.A.\\
		& 4 & 0.40 & 0.58 & 0.34 & 0.32 & 0.26 & 0.30 & 0.30 & 0.24\\ \hline
		
		\multirow{3}{*}{\ce{N2+} $\begin{dcases*} \\ \\ \end{dcases*}$} & 2 & n.c. & n.c. & n.c. & n.c. & n.c. & n.c. & n.c. & n.c.\\
		& 3 & --0.19 & --0.34 & --0.55 & N.A.        & --0.59 & --0.79 & --0.79 & N.A.\\
		& 4 & 0.75 & 1.09 & 0.62 & 0.70 & 0.49 & 0.64 & 0.64 & 0.47\\ \hline
		
		\multirow{3}{*}{\ce{BO} $\begin{dcases*} \\ \\ \end{dcases*}$} & 2 & 1.25 & 1.79 & 1.08 & 1.08 & 0.28 & 0.28 & 0.28 & 0.28\\
		& 3 & 0.09 & 0.06 & --0.20 & N.A.        & --0.29 & --0.42 & --0.42 &  N.A.\\
		& 4 & 0.50 & 0.84 & 0.38 & 0.36 & 0.26 & 0.31 & 0.31 & 0.24\\ \hline
		
		\multirow{3}{*}{\ce{CN} $\begin{dcases*} \\ \\ \end{dcases*}$} & 2 & n.c. & n.c. & n.c. & n.c. & 0.86 & 0.86 & 0.86 & 0.86\\
		& 3 & --0.17 & --0.38 & --0.58 & N.A.        & --0.58 & --0.83 & --0.83 & N.A.\\
		& 4 & 0.67 & 1.00 & 0.56 & 0.62 & 0.45 & 0.58 & 0.58 & 0.43\\ \hline
		
		\multirow{3}{*}{\ce{NF} $\begin{dcases*} \\ \\ \end{dcases*}$} & 2 & 1.67 & 2.06 & 1.45 & 1.45 & 0.68 & 0.68 & 0.68 & 0.68\\
		& 3 & --0.17 & --0.48 & --0.71 & N.A.        & --0.57 & --0.92 & --0.92 & N.A.\\
		& 4 & 0.36 & 0.64 & 0.23 & 0.23 & 0.15 & 0.20 & 0.20 & 0.12\\ \hline
		
		\multirow{3}{*}{\ce{NO} $\begin{dcases*} \\ \\ \end{dcases*}$} & 2 & n.c. & n.c. & n.c. & n.c. & 0.68 & 0.68 & 0.68 & 0.68\\
		& 3 & --0.12 & --0.35 & --0.51 & N.A.        & --0.45 & --0.70 & --0.70 & N.A.\\
		& 4 & 0.53 & 0.76 & 0.45 & 0.44 & 0.35 & 0.42 & 0.42 & 0.32\\ \hline
		
		\multirow{3}{*}{\ce{BF} $\begin{dcases*} \\ \\ \end{dcases*}$} & 2 & 0.51 & 0.62 & 0.45 & 0.45 & 0.05 & 0.05 & 0.05 & 0.05\\
		& 3 & 0.00 & --0.06 & --0.16 & N.A.        & --0.22 & --0.30 & --0.30 & N.A.\\
		& 4 & 0.15 & 0.24 & 0.10 & 0.13 & 0.09 & 0.11 & 0.11 & 0.07\\ \hline

		\multirow{3}{*}{\ce{F2} $\begin{dcases*} \\ \\ \end{dcases*}$} & 2 & n.c. & n.c. & n.c. & n.c. & n.c. & n.c. & n.c. & n.c.\\
		& 3 & --1.89 & --4.07 & --4.12 & N.A.        & --3.04 & --4.61 & --4.61 & N.A.\\
		& 4 & 0.67 & 0.21 & 0.11 & 2.70 & 0.77 & 1.38 & 1.38 & 1.06\\
	\end{tabular}
	
	\caption{Errors for equilibrium geometries of various diatomics in pm, as computed by variational theories with doubles truncated at degrees 2, 3, and 4, and triples truncated at degree 2, in the cc-pCVDZ basis set, relative to CCSDTQ(P). n.c. indicates that we could not converge to a minimum.} \label{tbl:diatomic_geom_t2}
\end{table}

\begin{table}
	\singlespacing
	\centering
	\begin{tabular}{cc|cccccccc}
		& & OVUCC & OVCC & OSCVCC & O$\lambda$ & OUDCT & OVDCT & OSVDCT & O$\lambda$DCT \\ \hline
		\multirow{3}{*}{\ce{N2} $\begin{dcases*} \\ \\ \end{dcases*}$} & 2 & --140 & --157 & --132 & --132 & --42 & --42 & --42 & --42\\
		& 3 & 21 & 39 & 52 & N.A.        & 49 & 68 & 68 & N.A.\\
		& 4 & --48 & --64 & --41 & --45 & --35 & --42 & -- 42 & --34\\ \hline
		
		\multirow{3}{*}{\ce{CO} $\begin{dcases*} \\ \\ \end{dcases*}$} & 2 & --84 & --108 & --73 & --73 & --28 & --28 & --28 & --28\\
		& 3 & --10 & --7 & 13 & N.A.        & 19 & 31 & 31& N.A.\\
		& 4 & --51 & --77 & --42 & --39 & --31 & --36 & --36 & --29\\ \hline
		
		\multirow{3}{*}{\ce{N2+} $\begin{dcases*} \\ \\ \end{dcases*}$} & 2 & n.c. & n.c. & n.c. & n.c. & n.c. & n.c. & n.c. & n.c.\\
		& 3 & 22 & 39 & 64 & N.A.        & 65 & 86 & 86 & N.A.\\
		& 4 & --88 & --137 & --73 & --85 & --55 & --80 & --80  & --54\\ \hline
		
		\multirow{3}{*}{\ce{BO} $\begin{dcases*} \\ \\ \end{dcases*}$} & 2 & --131 & --205 & --116 & --116 & --24 & --24 & --24 & --24\\
		& 3 & --11 & --14 & 12 & N.A.        & 22 & 31 & 31 & N.A.\\
		& 4 & --47 & --84 & --36 & --34 & --23 & --28 & --28 & --21\\ \hline
		
		\multirow{3}{*}{\ce{CN} $\begin{dcases*} \\ \\ \end{dcases*}$} & 2 & n.c. & n.c. & n.c. & n.c. & --107 & --107 & --107 & --107\\
		& 3 & 11 & 22 & 40 & N.A.        & 43 & 59 & 59 & N.A.\\
		& 4 & --64 & --100 & --55 & --60 & --42 & --58 & -- 58 & --41\\ \hline
		
		\multirow{3}{*}{\ce{NF} $\begin{dcases*} \\ \\ \end{dcases*}$} & 2 & --106 & --131 & --93 & --93 & --41 & --41 & --41 &  --41\\
		& 3 & 13 & 31 & 42 & N.A.        & 33 & 52 & 52 & N.A.\\
		& 4 & --15 & --31 & --8 & --8 & --3 & --6 & --6 & --1\\ \hline
		
		\multirow{3}{*}{\ce{NO} $\begin{dcases*} \\ \\ \end{dcases*}$} & 2 & n.c. & n.c. & n.c. & n.c. & --95 & --95 & --95 &  --95\\
		& 3 & 12 & 30 & 50 & N.A.        & 47 & 70 & 70 & N.A.\\
		& 4 & --62 & --96 & --54 & --51 & --39 & --50 & --50 & --34\\ \hline
		
		\multirow{3}{*}{\ce{BF} $\begin{dcases*} \\ \\ \end{dcases*}$} & 2 & --23 & --30 & --19 & --19 & --1 & --1 & --1 & --1\\
		& 3 & --2 & 0 & 6 & N.A.        & 9 & 13 & 13 & N.A.\\
		& 4 & --8 & --13 & --5 & --6 & --3 & --5 & --5 & --3\\ \hline
		
		\multirow{3}{*}{\ce{F2} $\begin{dcases*} \\ \\ \end{dcases*}$} & 2 & n.c. & n.c. & n.c. & n.c. & n.c. & n.c. & n.c. & n.c.\\
		& 3 & 75 & 168 & 169 & N.A.        & 124 & 188 & 188 & N.A.\\
		& 4 & --23 & 9 & 10 & --147 & --34 & --80 & --80 & --50\\
		
	\end{tabular}
	
	\caption{Errors for equilibrium harmonic vibrational frequencies of various diatomics in \cm, as computed by variational theories with doubles truncated at degrees 2, 3, and 4, and triples truncated at degree 2, in the cc-pCVDZ basis set, relative to CCSDTQ(P). The abbreviation n.c. means that we could not converge to a minimum.} \label{tbl:diatomic_freq_t2}
\end{table}

\begin{figure}
	\caption{The mean absolute errors and standard deviations of the signed errors in the (\protect\subref{subfig:dynamic-geom}) geometries and (\protect\subref{subfig:dynamic-freq}) frequencies of diatomics, relative to CCSDTQ(P), for approximate orbital-optimized methods with a degree two iterative triples correction and doubles terms truncated at degree 3 or 4, using the cc-pCVDZ basis set.}
	\label{fig:dynamic-graphics}
	\begin{subfigure}[b]{0.45\textwidth}
		\caption{}
		\label{subfig:dynamic-geom}
		\includegraphics[height=6.5cm]{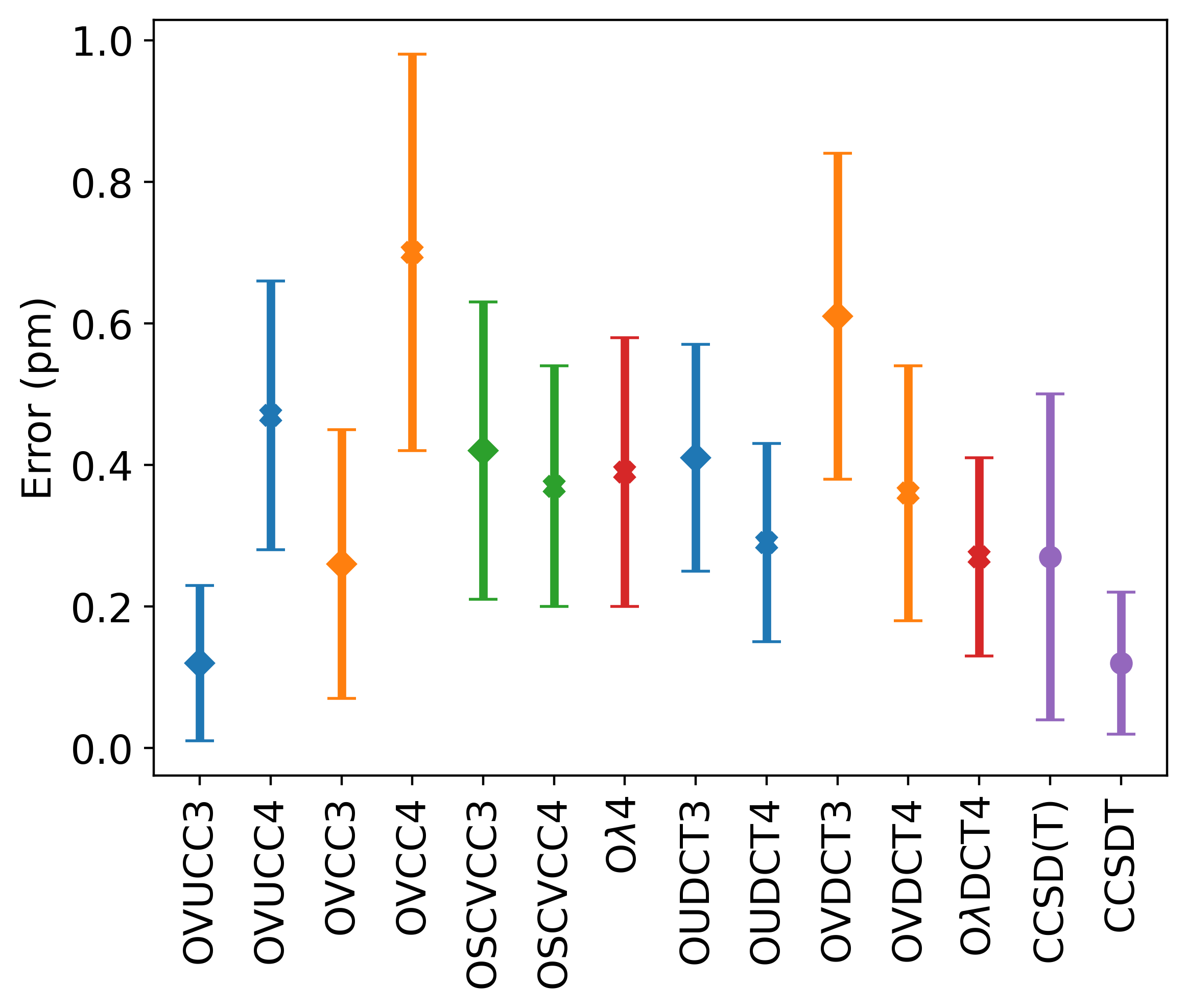}
	\end{subfigure}
	\begin{subfigure}[b]{0.45\textwidth}
		\caption{}
		\label{subfig:dynamic-freq}
		\includegraphics[height=6.5cm]{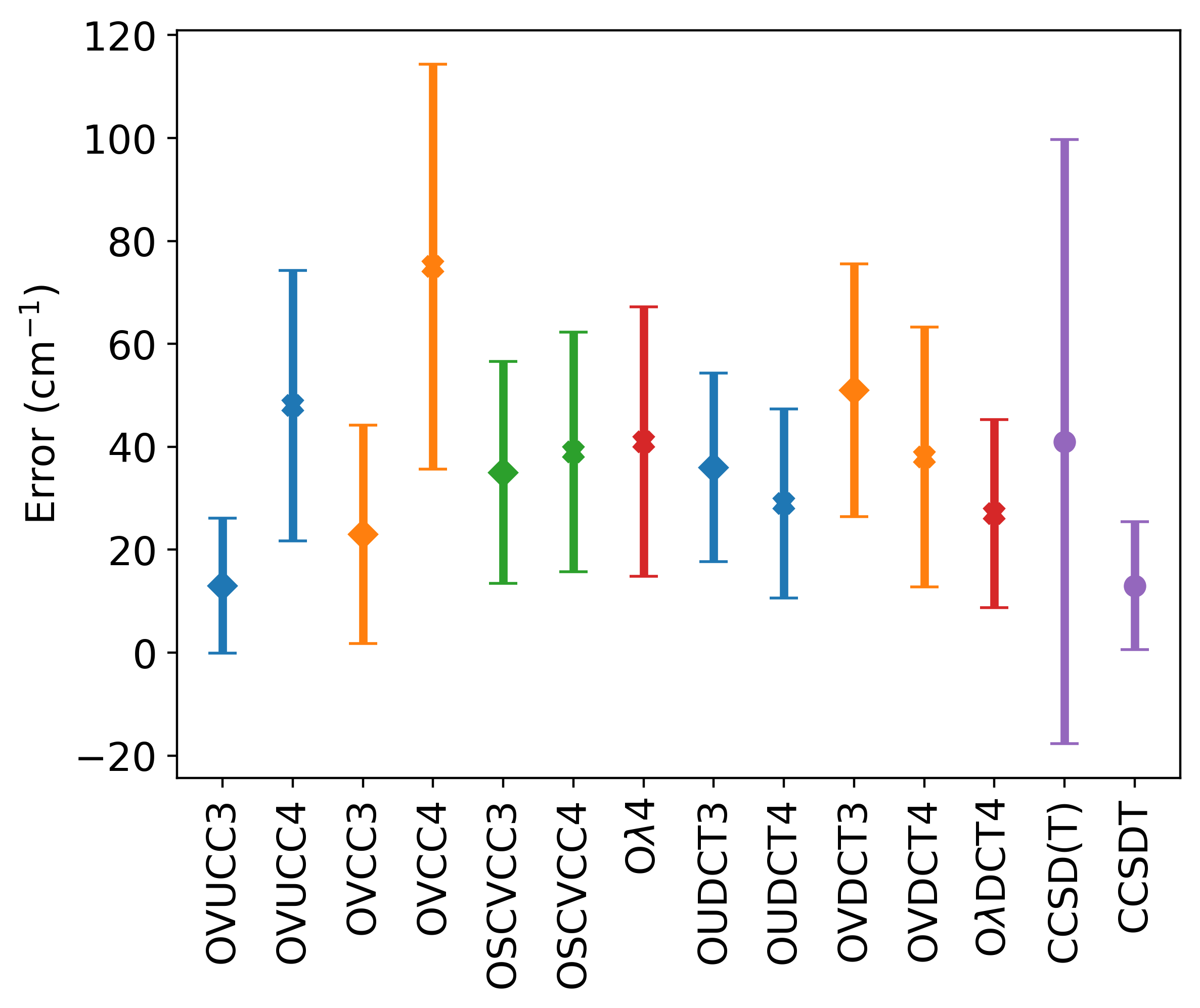}
	\end{subfigure}
\end{figure}

We observe that the degree two theories can encounter convergence difficulties, and the degree two variants not using DCT have especially large errors. To better analyze the remaining theories, we depict average statistics for equilibrium geometries and harmonic vibrational frequencies in Figure \ref{fig:dynamic-graphics}.

Among the orbital optimized reduced density parameterization theories, the best performance is surprisingly observed by OVUCC3+T2, with 0.12 pm equilibrium bond length error, and 13 \cm\ harmonic vibrational frequency error. We see a significant increase in the mean absolute error upon including the degree four terms, so this result is certainly not a consequence of being converged to the exact variational unitary coupled cluster theory. A perturbative explanation is not apparent to us, as the +T2 terms are expected to affect the energy first at $
\mathcal{O}(\mu^4)$, but the $
\mathcal{O}(\mu^4)$ doubles terms are excluded. A more careful analysis of how perturbation theory interacts with orbital optimization is likely needed. The VUCC3+T2 theory similarly has a good performance of 0.26 pm and 23 \cm\ errors, respectively, which is comparable to the performance of O$\lambda$DCT4+T2 and OUDCT4+T2. O$\lambda$DCT4+T2 has strictly fewer terms.

We now compare O$\lambda$DCT4+T2 to coupled cluster theories. We see that O$\lambda$DCT4+T2 is not competitive with the iterative triples correction of CCSDT but is competitive with the perturbative correction of CCSD(T). CCSD(T) has mean errors of 0.27 pm and 41 \cm, while CCSDT has mean errors of 0.12 pm and 13 \cm. This also means that the OVUCC3+T2 theory is competitive with CCSDT.

\subsubsection{Degree Three Triples}

\begin{figure}
	\caption{The mean absolute errors and standard deviations of the signed errors in the (\protect\subref{subfig:dynamic-geom3}) geometries and (\protect\subref{subfig:dynamic-freq3}) frequencies of diatomics, relative to CCSDTQ(P), for approximate orbital-optimized methods with a degree three iterative triples correction and doubles terms truncated at degree 3 or 4, using the cc-pCVDZ basis set. Theories that do not converge consistently are excluded.}
	\label{fig:dynamic-graphics3}
	\begin{subfigure}[b]{0.45\textwidth}
		\caption{}
		\label{subfig:dynamic-geom3}
		\includegraphics[height=6.5cm]{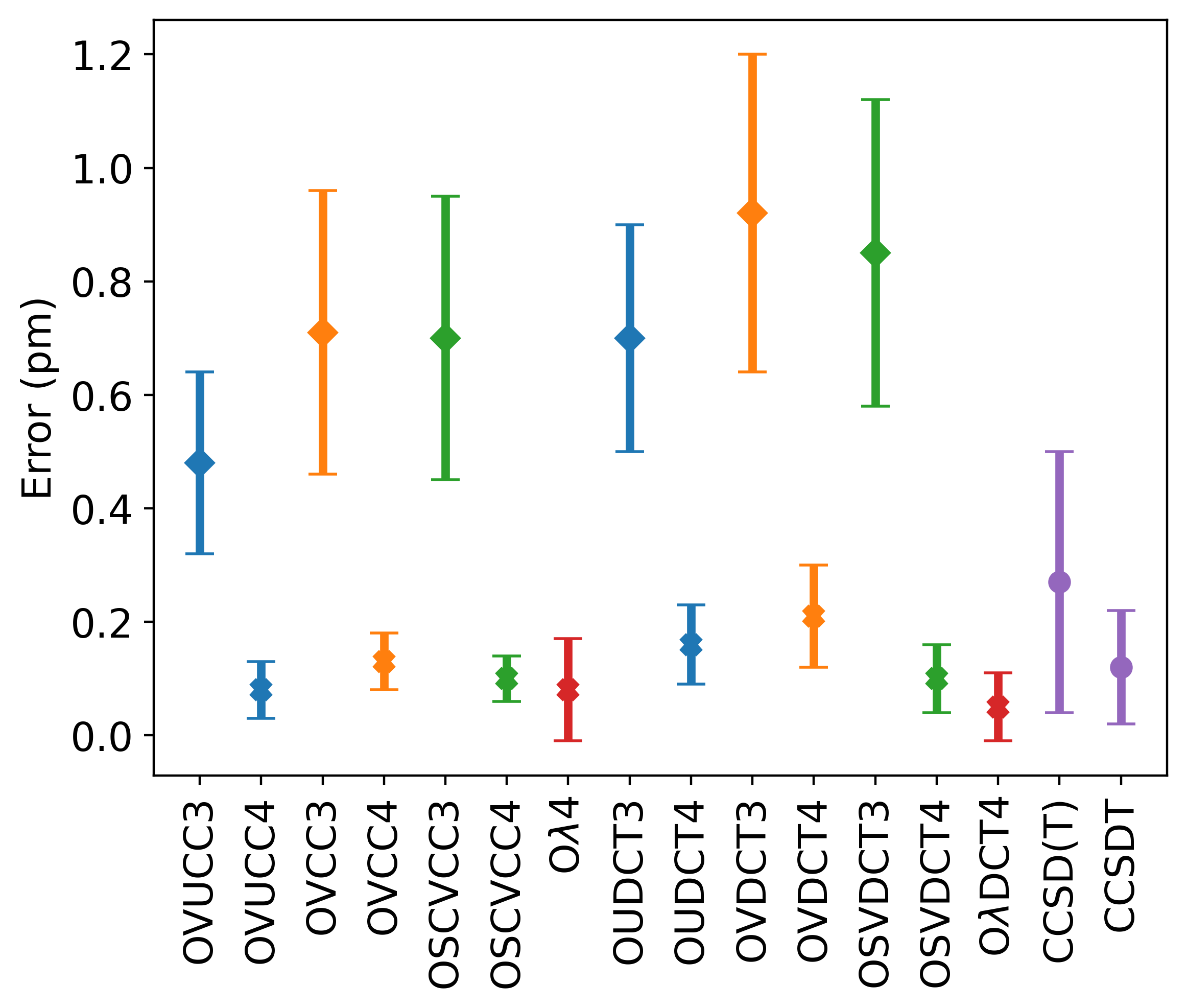}
	\end{subfigure}
	\begin{subfigure}[b]{0.45\textwidth}
		\caption{}
		\label{subfig:dynamic-freq3}
		\includegraphics[height=6.5cm]{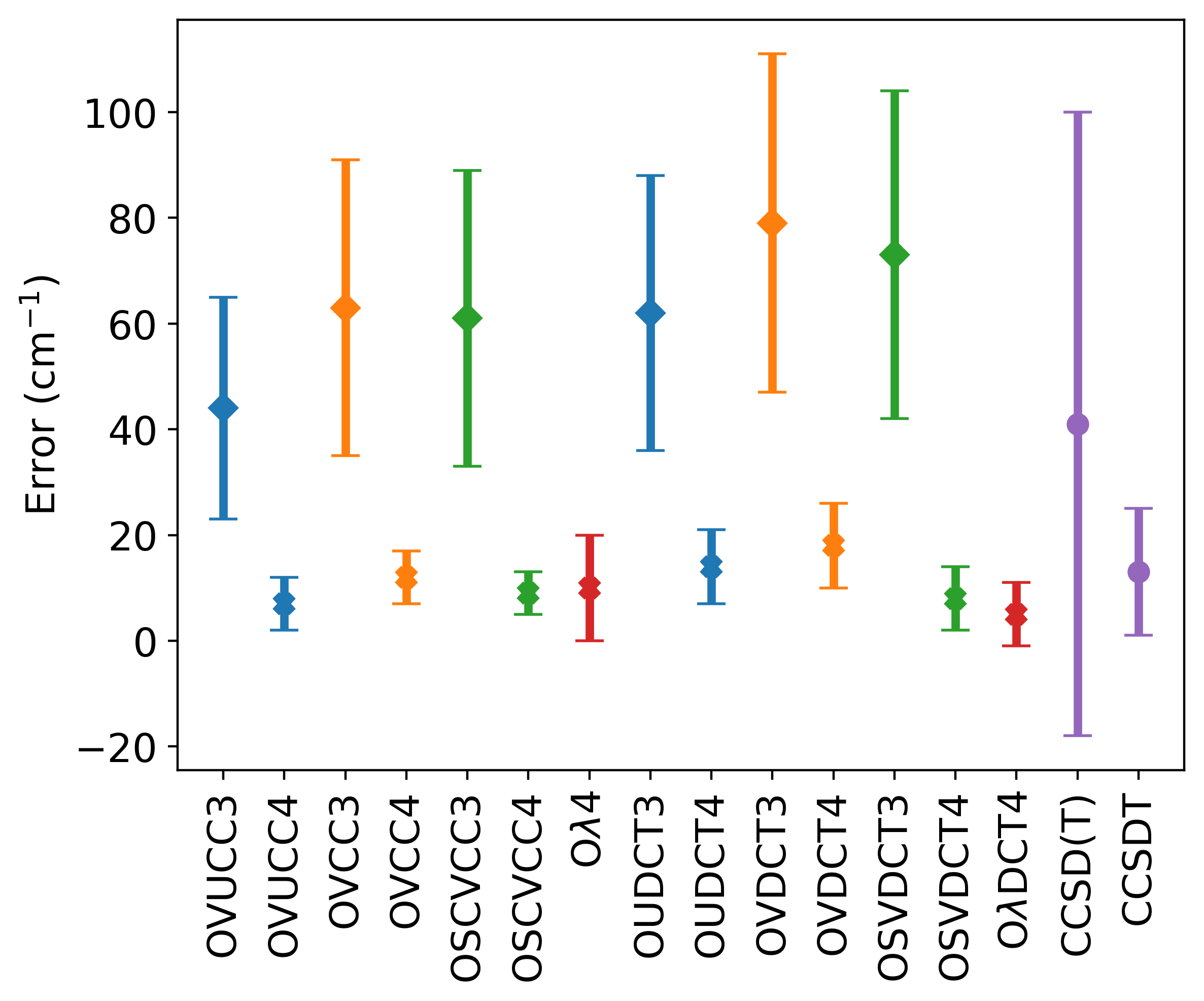}
	\end{subfigure}
\end{figure}

We now include the additional terms of the +T3 correction for each of the theories we consider. The errors in equilibrium bond lengths are given in Table \ref{tbl:diatomic_geom_t3}, and the errors in harmonic vibrational frequencies are given in Table \ref{tbl:diatomic_freq_t3}. We observe convegence failures with the O$\lambda$ and O$\lambda$DCT methods when degree four terms are excluded. (These are the only theories without degree three pure doubles terms.) The average performance of the remaining theories is shown in Figure \ref{fig:dynamic-graphics3}.

\begin{table}
	\singlespacing
	\centering
	\begin{tabular}{cc|cccccccc}
		& & OVUCC & OVCC & OSCVCC & O$\lambda$ & OUDCT & OVDCT & OSVDCT & O$\lambda$DCT \\ \hline
		\multirow{2}{*}{\ce{N2} $\begin{dcases*} \\ \\ \end{dcases*}$} & 3 & --0.43	& --0.65 &	--0.64 & 0.56 &	--0.61	& --0.81	& --0.76	& 0.11\\
		& 4 & --0.03 &	--0.06&	--0.04&	0.09&	--0.07&	--0.10&	--0.03&	0.03\\ \hline
		\multirow{2}{*}{\ce{CO} $\begin{dcases*} \\ \\ \end{dcases*}$} & 3 & --0.35	&--0.50	&--0.48& 0.28 &	--0.51&	--0.66&	--0.60 & 0.01\\
		& 4 & --0.09 & --0.11	&--0.09&	0.02&	--0.15&	--0.20&	--0.11&	--0.03\\ \hline
		
		\multirow{2}{*}{\ce{N2+} $\begin{dcases*} \\ \\ \end{dcases*}$} & 3 & --0.66 & --0.91 & --0.90 & 1.94 & --0.93 & --1.15 & --1.07 & n.c.\\
		& 4 &--0.03 & --0.10 & --0.07 & 0.21 & --0.13 & --0.16 & --0.03 & 0.07\\ \hline
		
		\multirow{2}{*}{\ce{BO} $\begin{dcases*} \\ \\ \end{dcases*}$} & 3 & --0.43 & --0.57 & --0.55 & 0.48 & --0.64 & --0.79 & --0.70 & --0.04\\
		& 4 &--0.15 & --0.19 & --0.15 & --0.01 & --0.25 & --0.31 & --0.17 & --0.08\\ \hline
		
		\multirow{2}{*}{\ce{CN} $\begin{dcases*} \\ \\ \end{dcases*}$} & 3 &--0.66 & --0.95 & --0.94 & n.c. & --0.93 & --1.20 & --1.12 & 0.40\\
		& 4 &--0.06 & --0.12 & --0.09 & 0.19 & --0.14 & --0.19 & --0.06 & 0.06\\ \hline
		
		\multirow{2}{*}{\ce{NF} $\begin{dcases*} \\ \\ \end{dcases*}$} & 3 & --0.57 & --0.98 & --0.97 & 1.22 & --0.85 & --1.22 & --1.14 & 0.52\\
		& 4 &--0.16 & --0.21 & --0.19 & 0.04 & --0.26 & --0.34 & --0.19 & --0.05\\ \hline
		
		\multirow{2}{*}{\ce{NO} $\begin{dcases*} \\ \\ \end{dcases*}$} & 3 & --0.56 & --0.84 & --0.83 & 1.54 & --0.78 & --1.04 & --0.97 & 0.35\\
		& 4 &--0.08 & --0.13 & --0.10 & 0.11 & --0.17 & --0.23 & --0.11 & 0.02\\ \hline
		
		\multirow{2}{*}{\ce{BF} $\begin{dcases*} \\ \\ \end{dcases*}$} & 3 & --0.19 & --0.30 & --0.29 & 0.27 & --0.37 & --0.46 & --0.42 & --0.08\\
		& 4 &--0.08 & --0.10 & --0.09 & --0.01 & --0.10 & --0.12 & --0.06 & --0.05\\ \hline

		\multirow{2}{*}{\ce{F2} $\begin{dcases*} \\ \\ \end{dcases*}$} & 3 & --2.37 & --4.42 & --4.42 & n.c. & --3.37 & --4.87 & --4.85 & n.c.\\
		& 4 &--0.18 & --0.80 & --0.79 & 1.92 & --0.07 & 0.00 & 0.04 & 0.60\\
	\end{tabular}
	
	\caption{Errors for equilibrium geometries of various diatomics in pm, as computed by variational theories with doubles truncated at degrees 3, and 4, and triples truncated at degree 3, in the cc-pCVDZ basis set, relative to CCSDTQ(P). The abbreviation n.c. means that we could not converge to a minimum.} \label{tbl:diatomic_geom_t3}
\end{table}

\begin{table}
	\singlespacing
	\centering
	\begin{tabular}{cc|cccccccc}
		& & OVUCC & OVCC & OSCVCC & O$\lambda$ & OUDCT & OVDCT & OSVDCT & O$\lambda$DCT \\ \hline
		\multirow{2}{*}{\ce{N2} $\begin{dcases*} \\ \\ \end{dcases*}$} & 3 & 56 & 79 & 78 & --85 & 75 & 97 & 91 & --14\\
		& 4 &3 & 7 & 6 & --13 & 9 & 12 & 3 & --5\\ \hline
		
		\multirow{2}{*}{\ce{CO} $\begin{dcases*} \\ \\ \end{dcases*}$} & 3 & 36 & 49 & 47 & --26 & 53 & 66 & 58 & 4\\
		& 4 &11 & 14 & 11 & --2 & 19 & 24 & 13 & 4\\ \hline
		
		\multirow{2}{*}{\ce{N2+} $\begin{dcases*} \\ \\ \end{dcases*}$} & 3 & 73 & 102 & 100 & --407 & 99 & 123 & 115 & n.c.\\
		& 4 &4 & 12 & 8 & --27 & 13 & 14 & 1 & --8\\ \hline
		
		\multirow{2}{*}{\ce{BO} $\begin{dcases*} \\ \\ \end{dcases*}$} & 3 & 37 & 47 & 45 & --48 & 53 & 64 & 56 & 5\\
		& 4 &16 & 19 & 15 & 2 & 24 & 29 & 17 & 9\\ \hline
		
		\multirow{2}{*}{\ce{CN} $\begin{dcases*} \\ \\ \end{dcases*}$} & 3 & 52 & 72 & 71 & n.c. & 71 & 90 & 83 & --45\\
		& 4 &3 & 9 & 6 & --20 & 9 & 12 & 2 & --7\\ \hline
		
		\multirow{2}{*}{\ce{NF} $\begin{dcases*} \\ \\ \end{dcases*}$} & 3 & 30 & 53 & 53 & --90 & 43 & 63 & 60 & --40\\
		& 4 &8 & 12 & 10 & --4 & 14 & 17 & 10 & 3\\ \hline
		
		\multirow{2}{*}{\ce{NO} $\begin{dcases*} \\ \\ \end{dcases*}$} & 3 & 62 & 88 & 86 & --405 & 83 & 108 & 100 & --52\\
		& 4 &11 & 17 & 13 & --13 & 21 & 28 & 13 & --1\\ \hline
		
		\multirow{2}{*}{\ce{BF} $\begin{dcases*} \\ \\ \end{dcases*}$} & 3 & 9 & 13 & 13 & --11 & 17 & 20 & 18 & 5\\
		& 4 &4 & 5 & 4 & 1 & 6 & 6 & 3 & 3\\ \hline
		
		\multirow{2}{*}{\ce{F2} $\begin{dcases*} \\ \\ \end{dcases*}$} & 3 & 93 & 179 & 179 & n.c. & 136 & 197 & 196 & n.c.\\
		& 4 &9 & 44 & 43 & --111 & --1 & --11 & --11 & --32\\
		
	\end{tabular}
	
	\caption{Errors for equilibrium harmonic vibrational frequencies of various diatomics in \cm, as computed by variational theories with doubles truncated at degrees 3, and 4, and triples truncated at degree 3, in the cc-pCVDZ basis set, relative to CCSDTQ(P). The abbreviation n.c. means that we could not converge to a minimum.} \label{tbl:diatomic_freq_t3}
\end{table}

Upon cross-referencing Figure \ref{fig:dynamic-graphics} with Figure \ref{fig:dynamic-graphics3}, it is apparent that starting from the theories with degree three doubles and degree two triples, the degree four doubles and degree three triples should both be included to see improvement.

Once these additional terms are included, all theories considered are superior to CCSD(T). The best performing theory is the O$\lambda$DCT4+T3 developed in this work, with mean absolute errors of 0.05 pm and 5 \cm, in comparison to the 0.12 pm and 13 \cm\ respective errors of CCSDT. This confirms the competitiveness of our method with coupled cluster and the alternative orbital-optimized reduced density matrix parameterizations considered here. A more definitive statement would require a larger test set.

We caution that our test set excluded \ce{F2}, for which O$\lambda$DCT4+T3 has a relatively large error of 0.60 pm and --32 \cm. While this is a large improvement on O$\lambda$ without DCT, other theories are more accurate. The cause of this is not obvious to us, but indicates that the performance of DCT is still not completely understood. We refer the reader to a recent study\cite{Zhang2021} on the roles of the cumulant and the rest of the density matrices in describing \ce{F2}.

\section{Conclusions}

In this study, we have proposed a new ansatz for density cumulant theory, in which the cumulant is derived from a wavefunction parameterization in terms of the particle-hole cumulants.\cite{Misiewicz2021:I} We combine this with orbital optimization, so cumulants of the form $\gamma^i_a$ are eliminated and the orbitals may always be chosen to be the natural orbitals. We call this ansatz O$\lambda$DCT.

Through formal analysis and the implementation and benchmarking of approximations of this parameterization, we have been able to answer the questions set out in the introduction:

\begin{itemize}
	\item Near-zero denominators coupling occupied and virtual orbitals are eliminated in the O$\lambda$DCT ansatz, as a consequence of $\gamma^i_a$ being rigorously zero. In highly correlated settings, near zero denominators arise in \eqref{eq:amplitude-gradient} when a natural orbital has an occupation number near 0.5, and these are not removed by O$\lambda$DCT. This is not curable by any choice of ansatz, as it would imply the natural orbital occupation numbers cannot vary.
	\item Methods of the ansatz are rigorously size-extensive. Unlike the previous OUDCT ansatz, in O$\lambda$DCT, both the parameterizations of the reduced density matrix and the derived amplitude residuals consist of connected tensors only.
	\item The failure of OUDCT methods for describing \ce{H2} dissociation was due to poor convergence of the cumulant partial trace with respect to unitary coupled cluster amplitudes. By contrast, this is perfectly converged at degree two within the O$\lambda$DCT ansatz and a minimal basis set. Numerical experiments on systems with larger basis sets confirm that O$\lambda$DCT uniformly improves on ODC-12's description of \ce{H2} dissociation. Combined with the good performance of O$\lambda$DCT4+T2 on \ce{Be2}, this suggests that the O$\lambda$DCT ansatz has regained the static correlation tolerance of ODC-12. More studies are needed to confirm this.
	\item O$\lambda$DCT theories have strictly fewer tensor contractions than OUDCT theories. At degree four in doubles, the number is reduced by about one half.
	\item For small systems, we observe that the amplitudes of approximate O$\lambda$DCT theories, including ODC-12, follow the theoretical exact amplitudes. This shows that our ansatz indeed ``explains'' the performance of O$\lambda$DCT, and supports our interpretation of ODC-12 in terms of this ansatz.
	\item Inclusion of more O$\lambda$DCT amplitudes of degree two does not \textit{on its own} improve the description of equilibrium properties for systems of more than two electrons. Rather, it is necessary to include triples. Inclusion of the degree two iterative triples terms gives results competitive with other orbital optimized reduced density matrix parameterizations as well as CCSD(T), but not CCSDT. Inclusion of the degree three terms gives a method superior to CCSDT, and at least in our test sets, the best performing method out of all the reduced density matrix parameterizations studied here.
\end{itemize}

Let us now remark on what this means for the future of density cumulant theory. First, we have presented a new ansatz, but our implementation here is based on deriving terms of the ansatz to low orders. This work has not presented a general form of the ansatz, which prohibits detailed analysis of it. We shall have much more to say on this topic in future work.\cite{Misiewicz2021:I} Second, one of the motivating factors for developing density cumulant theory is to generalize it to the multireference case. It seems likely to us that the ansatz to generalize will be the cumulant-based ansatz developed here.\cite{Misiewicz:2020p244102} If nothing else, the present article has demonstrated a new mechanism that compromises size-extensivity, which must be guarded against in any multi-configurational generalization.

In the single-reference setting, the O$\lambda$DCT4+T3 variant we have investigated seems worth implementing efficiently in an electronic structure package. Once this is done, we can perform more thorough benchmarks of the accuracy of the method, especially as regards thermochemistry and tolerance to static correlation. However, it has $\mathcal{O}(n^8)$ scaling, and an $\mathcal{O}(n^7)$ method analogous to CCSD(T) is also desirable. In this respect, it is worth studying the combination of the previously proposed perturbative triples correction\cite{Sokolov:2014p074111} with our O$\lambda$DCT4 theory. For an efficient gradient theory, it will be necessary to include this perturbative triples terms in the orbital optimization as well. Orbital optimization of a perturbative triples correction has been studied previously.\cite{doi:10.1063/1.4720382, doi:10.1063/5.0061351} Lastly, the original implementation of orbital optimized methods for dynamic correlation\cite{Sherrill:1998p4171} observed that it would be possible to orbital optimize the frozen core and frozen virtual spaces. In O$\lambda$DCT, the optimal orbitals can always be chosen as natural orbitals, so we expect these approximations to be exceptionally accurate.

\begin{acknowledgements}
	We acknowledge support from the National Science Foundation, Grant No. CHE-1661604 and CHE-2134792. We acknowledge helpful discusisons with Professors Alexander Yu. Sokolov, John F. Stanton, and Francesco A. Evangelista during earlier stages of this work.
\end{acknowledgements}

\section*{Data Availability Statement}
The code that support the findings of this study is openly available in GitHub, reference number \citenum{code}. All other data that supports the findings of this study are available within the article and its supplementary material.

\bibliography{abbreviations, JPMrefs.bib}
	
\end{document}